\documentclass[iop,apj]{emulateapj}
\usepackage{graphicx}
\usepackage{url}

\submitted{Accepted for publication in ApJ}
\journalinfo{Accepted for publication in ApJ}

\newcommand\kms{km~s$^{-1}$}

\newcommand\etal{{et~al.}}
\newcommand\mM{\ensuremath{(m{-}M)}}
\newcommand\feh{\ensuremath{[{\rm Fe/H}]}}

\newcommand\gacs{\ensuremath{g_{475}}}
\newcommand\Hwfc{\ensuremath{H_{160}}}
\newcommand\Iacs{\ensuremath{I_{814}}}
\newcommand\zacs{\ensuremath{z_{850}}}
\newcommand\MIacs{\ensuremath{M_{814}}}
\newcommand\Ipeng{\ensuremath{I_{\rm Peng}}}

\newcommand{\gIacs}{\ensuremath{g_{475}{-}I_{814}}}
\newcommand{\IHacs}{\ensuremath{I_{814}{-}H_{160}}}
\newcommand{\IHhst}{\ensuremath{I_{814}{-}H_{160}}}
\newcommand{\gzacs}{\ensuremath{g_{475}{-}z_{850}}}
\newcommand{\gIpeng}{\ensuremath{({g_{475}{-}I_{814}})_{\rm Peng}}}

\newcommand\magauto{{\tt MAG\_AUTO}}

\newcommand\magerriso{{\tt MAGERR\_ISO}}
\newcommand\fwhm{{\tt FWHM}}
\newcommand\classstar{{\tt CLASS\_STAR}}
\newcommand\kurt{\hbox{\it kurt}}
\newcommand\HST{{\it HST}}
\newcommand\hst{{\it HST}}
\newcommand\lta{\lesssim}
\newcommand\gta{\gtrsim}

\newcommand\Ref{\ensuremath{R_{\rm e}}}
\newcommand\sersic{S\'{e}rsic}

\shortauthors{Cho et al.}
\shorttitle{NGC 4874 Globular Clusters}

\begin{document}

\title{The Globular Cluster System of the Coma \lowercase{c}D Galaxy NGC\,4874 from \textit{Hubble Space Telescope} ACS and WFC3/IR Imaging\altaffilmark{*}}
\altaffiltext{*}{Based on observations with the NASA/ESA \textit{Hubble Space Telescope}, obtained from the Space Telescope Science Institute (STScI), which is operated by AURA, Inc., under NASA contract NAS 5-26555.
These observations are associated with program \#11712.}

\author{Hyejeon Cho\altaffilmark{1}, 
	John P.~Blakeslee\altaffilmark{2}, 
	Ana L.~Chies-Santos\altaffilmark{3,4}, 
	M.~James Jee\altaffilmark{1},
	Joseph B.~Jensen\altaffilmark{5},\\
	Eric W.~Peng\altaffilmark{6,7}
\& 	Young-Wook Lee\altaffilmark{1}
}
\altaffiltext{1}{Department of Astronomy and Center for Galaxy Evolution Research, 
Yonsei University, Seoul 03722, Korea; \texttt{hyejeon@yonsei.ac.kr; ywlee2@yonsei.ac.kr}}
\altaffiltext{2}{NRC Herzberg Astronomy \& Astrophysics, Victoria, BC V9E\,2E7, Canada; \texttt{john.blakeslee@nrc-cnrc.gc.ca}}
\altaffiltext{3}{Departamento de Astronomia, Instituto de F\'{\i}sica, UFRGS, Porto Alegre, R.S. 91501-970, Brazil}
\altaffiltext{4}{Departamento de Astronomia, IAGCA, Universidade de S\~{a}o Paulo, 05508-900 S\~{a}o Paulo, SP, Brazil}
\altaffiltext{5}{Department of Physics, Utah Valley University, Orem, Utah 84058, USA}
\altaffiltext{6}{Department of Astronomy, Peking University, Beijing 100871, China}
\altaffiltext{7}{Kavli Institute for Astronomy and Astrophysics, Peking University, Beijing 100871, China}

\begin{abstract}
   We present new \textit{Hubble Space Telescope} (\hst\,) optical and near-infrared
  (NIR) photometry of the rich globular cluster (GC) system of NGC\,4874, the cD
  galaxy in the core of the Coma cluster (Abell~1656). NGC\,4874 was observed
  with the \HST\ Advanced Camera for Surveys in the F475W (\gacs) and F814W
  (\Iacs) passbands and the Wide Field Camera~3 IR Channel in F160W (\Hwfc).
  The GCs in this field exhibit a bimodal optical color distribution with more than
  half of the GCs falling on the red side at\ $\gIacs>1$. Bimodality is also present,
  though less conspicuously, in the optical-NIR \IHacs\ color. 
  Consistent with past work, we find evidence for
  nonlinearity in the \gIacs\ versus \IHacs\ color-color relation. Our results thus
  underscore the need for understanding the detailed form of the color-metallicity
  relations in interpreting observational data on GC bimodality. We also find a very
  strong color-magnitude trend, or ``blue tilt,'' for the blue component of the optical
  color distribution of the NGC\,4874 GC system. A similarly strong trend is present
  for the overall mean \IHacs\ color as a function of magnitude; for $\MIacs<-10$~mag,
  these trends imply a steep mass-metallicity scaling with
  $Z\propto M_{\rm GC}^{1.4\pm0.4}$, but the scaling is not a simple power law 
  and becomes much weaker at lower masses. 
  As in other similar systems, the spatial distribution of the blue GCs is more extended 
  than that of the red GCs, partly because of blue GCs associated with surrounding cluster
  galaxies. In addition, the center of the GC system is displaced by $4\pm1$~kpc
  towards the southwest from the luminosity center of NGC\,4874, in the direction of
  NGC\,4872. 
  Finally, we remark on a dwarf elliptical galaxy with a noticeably asymmetrical GC distribution.
  Interestingly, this dwarf has a velocity of nearly $-$3000~\kms\ with respect to
  NGC\,4874; we suggest it is on its first infall into the cluster core and is
  undergoing stripping of its GC system by the cluster potential.
\end{abstract}


\section{Introduction}
\label{sec:intro}

All large galaxies possess globular cluster (GCs) populations, or systems, comprising
hundreds or thousands of individual GCs.
They are often used as discrete tracers of galaxy assembly, especially at the outer 
regions of large galaxies where they can be more easily observed than the faint 
integrated galaxy light. Moreover, they provide information on the various processes 
and progenitors that are present in the different phases of the build up of 
early-type galaxies (Zaritsky \etal\ 2014), which are the dominant structures 
at the centers of galaxy clusters (Dressler 1980).
Interestingly, the number of GCs, or the total mass of the GC
system, appears to be related to the mass of the dark matter halo of the galaxy
or galaxy cluster (Blakeslee \etal\ 1997; Blakeslee 1999; Bekki \etal\ 2008;
Spitler \& Forbes 2009; Alamo-Mart\'{\i}nez \etal\ 2013; Hudson \etal\ 2014).
However, even at a given mass, there is significant galaxy-to-galaxy scatter in
the number and other properties of the GC systems, and this scatter is likely
the result of 
environmental effects and stochastic variations in the galaxy formation histories
(e.g., Peng \etal\ 2008; Harris \etal\ 2013).
GC formation requires very high star formation rates, and 
thus the major star formation episodes and assembly histories of early-type
galaxies can be traced by the observed properties of their GC systems, such as
their colors, metallicities, and spatial distributions
(e.g., Brodie \& Strader 2006; Peng \etal\ 2006; Forte \etal\ 2014).

Since nearly all the GCs surrounding massive galaxies are old, metallicity is the
main stellar population variable among individual GCs within a GC system.
However, at present the acquisition of large samples of spectroscopic
metallicities for large numbers of GC systems is impractical.
With the exception of a few nearby early-type galaxies such as NGC\,5128 
(e.g., Beasley \etal\ 2008; Woodley \etal\ 2010),
the majority of the Lick index-type spectroscopic studies using 10\,m class 
telescopes have been carried out on samples of $\sim20$ to 50~GCs. 
One~recent major effort on this front is the SLUGGS survey (Brodie \etal\ 2014),
which uses the calcium II triplet (CaT) index  
as a metallicity proxy for over 1000 GCs within a sample of 25 GC systems.
CaT index distributions have been investigated by Usher \etal\ (2012), and 
a clear case of bimodality was presented by Brodie \etal\ (2012) for the 
edge-on S0 galaxy NGC\,3115.

Obtaining spectra with sufficiently high signal-to-noise ratios
to measure accurate metallicities for large samples of extragalactic GCs is
difficult; obtaining large samples of high-quality photometric colors is 
much simpler.
Optical color distributions for GC systems of massive galaxies 
are generally found to be bimodal (e.g., Peng \etal\ 2006);
however, this bimodality does not always hold when considering other color
combinations including near-infrared (NIR; Blakeslee \etal\ 2012; Chies-Santos
\etal\ 2012) or UV bands (Yoon \etal\ 2011a,b). Cantiello \& Blakeslee (2007; see
also Puzia \etal\ 2002) have shown that optical/NIR colors are better metallicity proxies than
purely optical colors. Optical wavelengths in old stellar systems are sensitive to the
red giant branch, but also to stars on the horizontal branch (HB) and near the main
sequence turn-off point, and therefore are degenerate in age and metallicity.
The NIR wavelength range is dominated by red giant branch stars, so the optical-NIR
colors are therefore mainly sensitive to metallicity.
Moreover, Yoon \etal\ (2006) have shown that non-linear
color-metallicity relations, possibly related to a sharp transition in the HB
morphology at a certain metallicity, can transform a unimodal metallicity distribution
into bimodal optical color distributions (see also Richtler 2006). Despite bimodal
metallicity distributions found for galaxies such as NGC\,3115 and the Milky Way, color-color
nonlinearities are present to a certain degree (Cantiello \etal\ 2014; Vanderbeke et al. 2014)
and are not yet fully understood. In a study of stacked GC spectra around the CaT region
for 10 galaxies, Usher \etal\ (2015) found galaxy-to-galaxy variations in the
CaT-color relations, implying that different types of galaxies require different
color-metallicity transformations for estimating GC metallicities from photometric data.

The situation is even more complex.
The bimodality in the optical color distributions of GCs around massive galaxies varies
in both the relative proportions of blue and red GCs and in the mean colors of these two
components. Even within a given galaxy there are variations.
For instance, the blue GCs of certain massive galaxies are observed to have redder colors 
at brighter magnitudes. This color-luminosity relation, or ``blue tilt'' 
(Harris \etal\ 2006; Mieske \etal\ 2006; Strader \etal\ 2006; Wehner \etal\ 2008)
has been suggested to be due to self-enrichment (e.g., Strader \& Smith 2008; 
Bailin \& Harris 2009). The blue tilt only becomes 
significant for GCs with masses above $\sim10^6 \, M_{\odot}$ but this effect has important 
implications for color distribution studies as the location of the blue and red peaks 
will vary with the magnitude range of the GCs considered.

The Coma cluster of galaxies is a truly massive and rich galaxy cluster at a mean redshift
of $z=0.024$ (Colless \& Dunn 1996), corresponding to a distance of about 100\,Mpc (for $h=0.7$).
Its virial mass of $2.7\times10^{15} \, M_{\odot}$ (Kubo \etal\ 2007) is roughly 
four times more massive than the Virgo cluster (see Carter \etal\ 2008; 
Durrell \etal\ 2014). As the anchor of comparison for studying properties of both
galaxies and clusters between the nearby and distant universe, Coma presents
an attractive opportunity for detailed studies of GCs in a dense cluster environment
(e.g., Harris 1987, Harris et al.\ 2009; Blakeslee \& Tonry 1995; Blakeslee \etal\ 1997;
Mar\'{\i}n-Franch \& Aparicio 2002). Analyzing the data from the \hst/ACS 
Coma Cluster Treasury Survey (hereafter ACSCCS; Carter \etal\ 2008),
Peng \etal\ (2011) discovered a population of intracluster globular clusters (IGCs)
in Coma that did not appear to be associated with any galaxy. 
This Coma IGC population was estimated to make up 
$\sim\,$30--45\% of all GCs in the cluster core and presents a bimodal color 
distribution with blue GCs greatly outnumbering red ones. Much of the remaining 
portion of Coma's core GC system belongs to its cD galaxy NGC\,4874, 
which also has a bimodal color distribution, but with a blue population that is 
somewhat redder than the blue IGCs.

This paper presents an analysis of new, significantly deeper ACS optical data than was obtained
by the ACSCCS, with the addition of new high resolution \hst/WFC3 NIR photometry of the
GC system surrounding NGC\,4874. 
We study the optical and optical-NIR color distributions, the nonlinear behavior in the
color--color relations as well as color--magnitude trends in the ACS/WFC F475W,
F814W and WFC3/IR F160W bandpass combinations. The spatial distribution of the GC
system is also explored within the wider ACS/WFC field of view. For consistency
with the ACSCCS studies (e.g., Carter \etal\ 2008; Peng \etal\ 2011) we adopt throughout
this paper a distance of 100~Mpc to Coma, giving a distance modulus of
$\mM=35.0$~mag. At this distance, 1\arcsec\ corresponds to a physical scale of
0.48~kpc.

\section{Observational Data Sets}
\label{sec:obs}

As part of \hst\ program GO-11711, we imaged NGC\,4874 with the Advanced Camera
for Surveys Wide Field Channel (ACS/WFC) for four orbits in F814W (\Iacs) and
one orbit in F475W (\gacs); to this, we added additional imaging in \gacs\
from GO-10861. The field of view of the ACS/WFC is approximately
$3\farcm37{\,\times\,}3\farcm37$.
The exposures were dithered to improve bad pixel rejection and to
fill in the 2\farcs5 gap between the two ACS/WFC detectors.
Following the standard pipeline processing at the Space
Telescope Science Institute's Mikulski Archive for Space Telescopes (MAST), we
used the stand-alone version of the empirical pixel-based charge-transfer
efficiency (CTE) correction algorithm of Anderson \& Bedin (2010) on each of the
individual calibrated ``flt'' exposures to remove the CTE trails from the ACS
data. The calibrated, CTE-corrected exposures were then processed with Apsis
(Blakeslee \etal\ 2003) to produce geometrically corrected, cosmic-ray rejected
stacked images with a final pixel scale of 0\farcs05~pix$^{-1}$. 

\begin{figure*}
\includegraphics[width=1.00\linewidth]{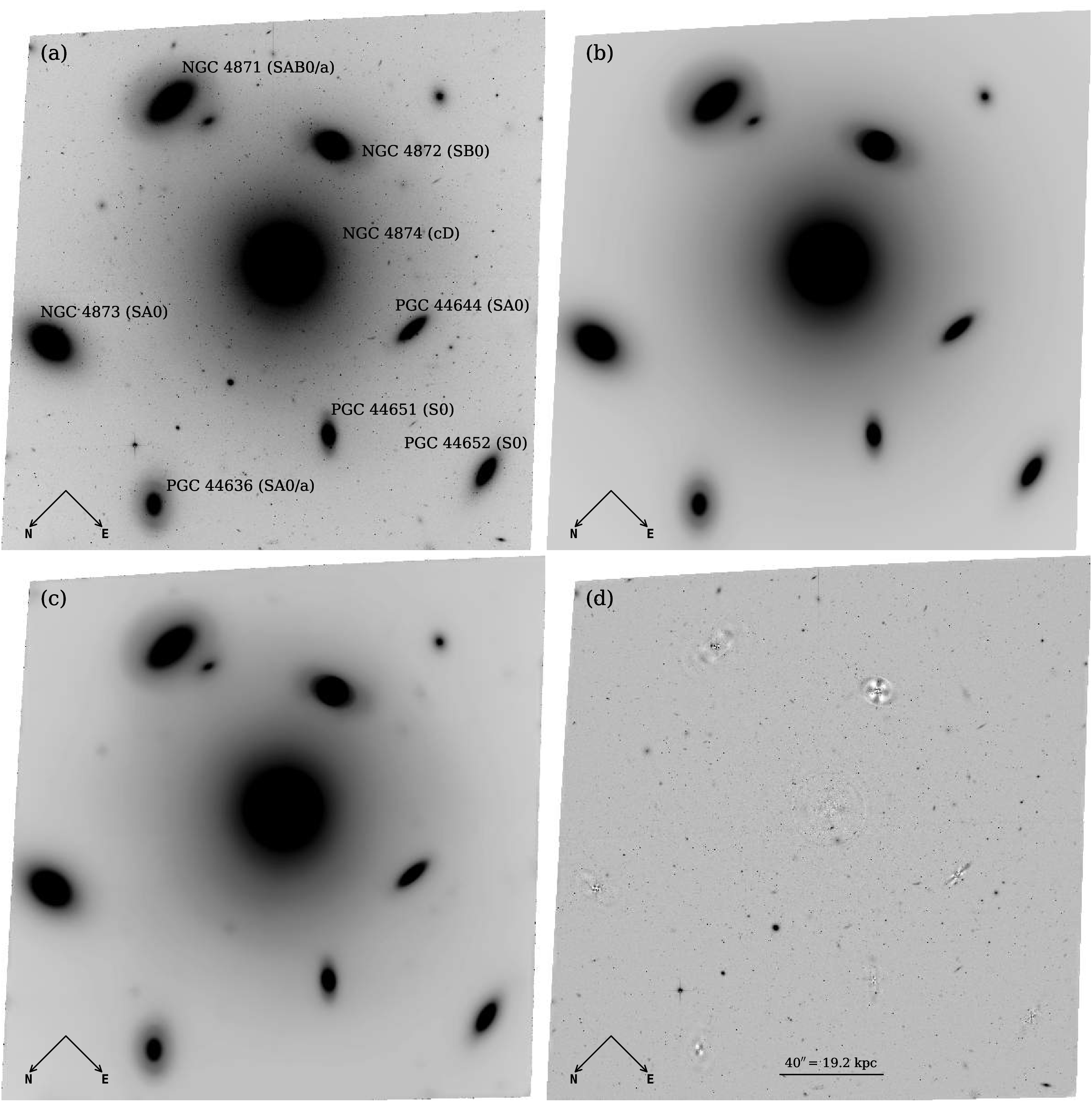}
\caption{
Stacked \HST\ ACS/WFC F814W image of the field roughly centered on NGC\,4874. The size of
the field is $3\farcm52\times3\farcm56$. (a)~The drizzled science image, shown at the
observed orientation. Bright extended galaxies in the field are labeled and 
their morphological types from NED are shown in parentheses. 
(b)~The sum of {\tt ELLIPSE}-generated 
isophotal models for NGC\,4874 and nine surrounding galaxies. 
(c)~The sum of galaxy isophotal models and a large-scale residual map constructed using
SExtractor. (d)~The final ``residual image'' used for the object detection with the
isophotal models and SExtractor background subtracted.
\label{fig:imgf814w}}
\end{figure*}

Figure~\ref{fig:imgf814w}\,(a) shows our ACS/WFC F814W image of the NGC\,4874 field,
along with the designations and morphological classifications from the NASA/IPAC
Extragalactic Database (NED)\footnote[8]{{\url{http://ned.ipac.caltech.edu}}} for
eight bright galaxies (including NGC\,4874 itself).

\begin{deluxetable*}{cclcrrc}
\tablecolumns{7}
\tablewidth{0pc}
\tablecaption{Observational Details of the Data Sets}
\tablehead{
\colhead{Program} & \colhead{Dataset} & \colhead{Instrument/} & \colhead{Bandpass} & \colhead{Exp. Time} &
\colhead{$m_{1}$\tablenotemark{a}}    & \colhead{Magnitude}\\
\colhead{ID}       & \colhead{}       & \colhead{Detector}    & \colhead{}         & \colhead{(sec)} &
\colhead{(mag)}                       & \colhead{Symbol}
}
\startdata
11711 & JB2I01010 & ACS/WFC & F475W &  2394.0 & 26.056 & $g_{475}$ \\
10861 & J9TY19040 & ACS/WFC & F475W &  2677.0 & 26.045 & $g_{475}$ \\
11711 & JB2I01020 & ACS/WFC & F814W & 10425.0 & 25.947 & $I_{814}$ \\
11711 & IB2I02040 & WFC3/IR & F160W & 10790.8 & 25.946 & $H_{160}$
\enddata
\tablenotetext{a}{Photometric zeropoints represent the magnitudes on the AB system corresponding to one count per second.}
\label{tab:obstab1}
\end{deluxetable*}

We also observed NGC\,4874 
with the Wide Field Camera~3 IR Channel (WFC3/IR)
in parallel for six additional orbits of GO-11711, with four of the orbits in the
longest wavelength F160W (\Hwfc) bandpass, during primary ACS/WFC observations
of the neighboring Coma giant elliptical NGC\,4889. The WFC3/IR focal plane
array consists of a single detector with a field of view of
$2\farcm27{\,\times\,}2\farcm05$. The calibrated WFC3/IR
\Hwfc\ exposures were retrieved from STScI/MAST and combined into a final
geometrically corrected image using the MultiDrizzle (Koekemoer \etal\ 2003; 
Fruchter \etal\ 2009) task in the PyRAF/STSDAS package\footnote{PyRAF and STSDAS 
are products of the Space Telescope Science Institute, operated by AURA 
for NASA.}. As in Blakeslee \etal\ (2012), we used an output pixel scale of 
0\farcs1~pix$^{-1}$, which is twice that of ACS/WFC.
Table~\ref{tab:obstab1} summarizes the observational details of our imaging
data; note that the two sets of F475W data from the two different programs
were combined by Apsis into a single stacked image. 

We corrected for Galactic extinction toward NGC\,4874 assuming $E(B{-}V)=0.0091$~mag 
(Schlegel \etal\ 1998) and the revised ACS/WFC and WFC3/IR extinction 
coefficients (for $R_V{\,=\,}3.1$) from Schlafly \& Finkbeiner (2011); 
the resulting corrections were small, amounting to 0.030, 0.014, and 0.005~mag 
in \gacs, \Iacs, and \Hwfc, respectively. When we derived K-corrections for 
12 Gyr model spectral energy distributions, which are redshifted to 
the NGC\,4874 distance (Ben\'{\i}tez 2000), with \feh\ $=-1.7$ and $-0.7$ 
(Bruzual \& Charlot 2003; C. Chung, private communication), corresponding to 
blue and red peak GCs, the average corrections are 0.05, 0.00, and $-0.02$~mag 
for \gacs, \Iacs, and \Hwfc, respectively. Since it is uncertain how good 
the evolutionary stellar population synthesis models are at NIR wavelengths 
and the estimated K-corrections are small but model-dependent, 
we have not applied them to our magnitudes and colors for GC candidates. 

In this paper, we calibrate the ACS photometry to the AB system following 
Bohlin (2012) and adopting the time-variable zero points from the online ACS Zeropoints 
Calculator\footnote{{\url{http://www.stsci.edu/hst/acs/analysis/zeropoints/zpt.py}}}. 
The WFC3 photometry is calibrated using the AB zero points from the online 
WFC3 zero~point tables\footnote{{\url{http://www.stsci.edu/hst/wfc3/phot\_zp\_lbn}}} 
(06~March 2012 revision). For reference, the adopted zero~points are provided 
in Table~\ref{tab:obstab1}; in the case of F475W, we used an exposure 
time-weighted average of the zero~points for the two different observations.

\section{Photometric Analysis}
\label{sec:phot}

\subsection{Galaxy and Background Subtraction}
\label{subsec:galbaksub}

In order to detect point-like objects embedded in the extended galaxy halo light, 
we first removed the smooth galaxy light profiles from the final combined images. 
We constructed elliptical isophotal models for each of the bright galaxies in each 
of the stacked bandpass images using the IRAF/STSDAS tasks {\tt ELLIPSE} and 
{\tt BMODEL}, which use the fitting algorithm and the uncertainty estimation 
method described by Jedrzejewski (1987) and Busko (1996).
We started by making an initial model (improved with later iterations)
of the brightest galaxy (NGC\,4874), then progressed by modeling the other
galaxies in order of their luminosity.  
When running {\tt ELLIPSE}, we first masked bright foreground stars, bad 
pixels, and any bright galaxies in the field except for the galaxy being fitted;
then we modeled the isophotes of the galaxy light distribution. 
Using the isophotal parameters from {\tt ELLIPSE}, we then
build a smooth galaxy model with {\tt BMODEL}.
After subtracting the model from the original image, we fitted isophotes of the
next brightest galaxy and subtracted this isophotal model as well. 
We repeated this process until we had subtracted ten galaxies in the ACS/WFC
image and four galaxies (NGC\,4874 itself and three surrounding galaxies) in the
WFC3/IR image.  As mentioned above,
it was necessary to model the galaxies iteratively in order to
achieve the cleanest model subtractions (e.g., Alamo-Mart\'{\i}nez \etal\ 2013).

\begin{figure*}
\includegraphics[width=1.0\linewidth]{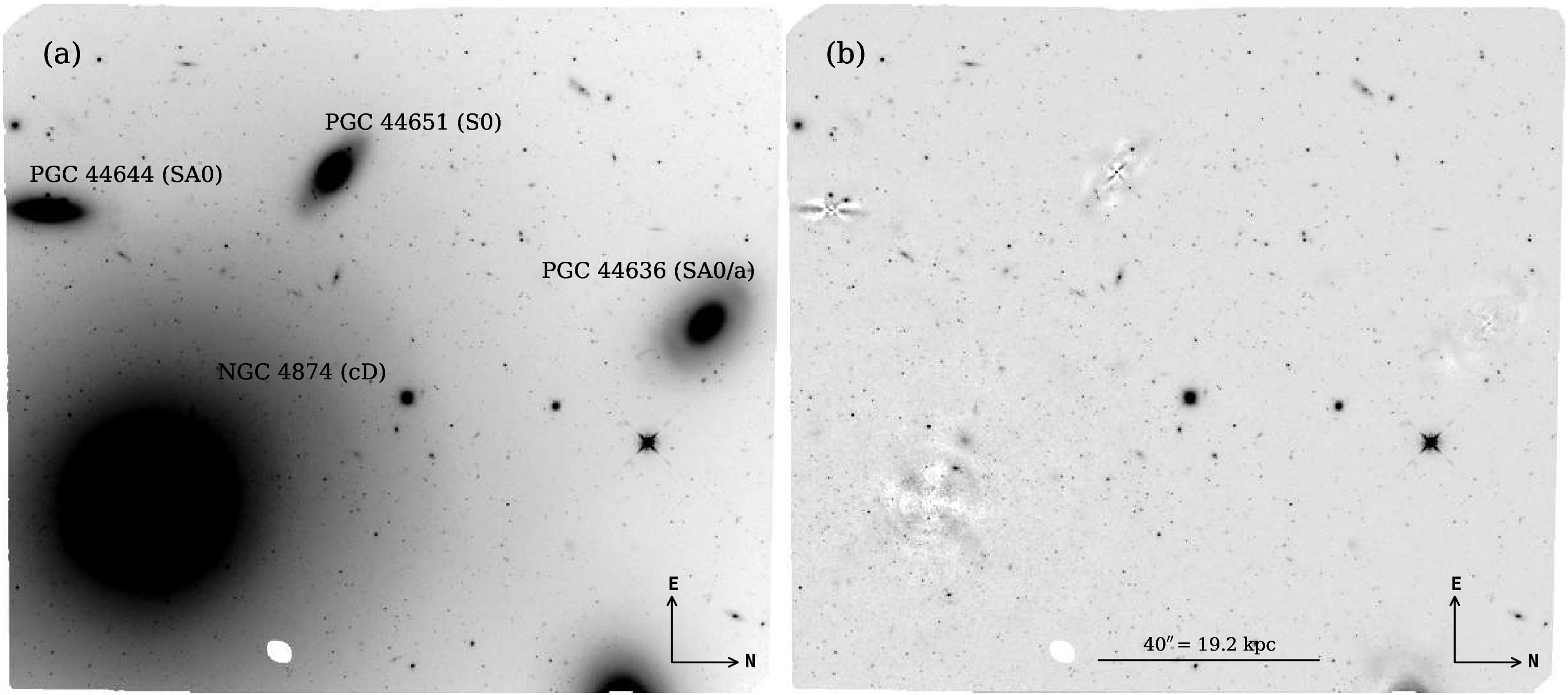}
\caption{Stacked \HST\ WFC3/IR F160W observation of NGC\,4874 and surrounding region.
The field size is $2\farcm36\times2\farcm09$; north is to the right and east is up. 
(a) The drizzled F160W science image; NGC\,4874 and three neighboring
galaxies are labeled. (b) The same image after subtracting our isophotal models of
the four large galaxies and a SExtractor-generated background map. 
\label{fig:imgf160w}}
\end{figure*}

After subtracting the elliptical isophotal models, we modeled the residual 
background using SExtractor (Bertin \& Arnouts 1996) to fit a two-dimensional 
bicubic spline with the parameters ${\tt BACK\_SIZE} = 32$ and 
${\tt BACK\_FILTERSIZE} = 3$. 
This removes residual structure on scales much larger than the 
full width half maximum (FWHM) of the point spread function (PSF), 
and thus does not detrimentally affect the point source photometry (see 
Jord\'{a}n \etal\ 2004). 
We note that subtraction of the isophotal model generated by the
{\tt BMODEL} task sometimes results in a noticeable discontinuity in surface
brightness at the ``edge'' of the model. However, because we modeled the
galaxies to very low  surface brightness levels, and performed careful
iterative modeling to achieve flat local background levels,
such residual ``edge'' features were generally in the noise.
In addition, spurious detections associated with the model edges would 
be removed by our point source selection criteria described below.
Panels (b) through (d) of Figure~\ref{fig:imgf814w} respectively show our
combined isophotal models for the galaxies labeled in panel (a) plus two
additional galaxies; the isophotal galaxy models plus the residual background map;
and the final ``residual image'' after subtracting the galaxy and residual background models.
For comparison, the stacked WFC3/IR F160W science image and residual image
following galaxy and background map subtraction are presented in 
Figure~\ref{fig:imgf160w}. Because of the smaller field of view, only NGC\,4874 
and the three other galaxies (labeled) were modeled. Disky residuals 
are noticeable in some cases, but the subtracted images are 
generally quite clean, revealing many faint sources.

\subsection{Object Detection and GC Candidate Selection}
\label{subsec:detection}

Object detection and photometric measurements were performed on the 
final residual images using SExtractor independently for each bandpass
(i.e., in ``single-image mode'').
For the ACS photometry, we used the RMS weight images produced by Apsis
as the SExtractor weight images (type {\tt MAP\_RMS}).
For the WFC3/IR F160W photometry, we used a variance map (type {\tt MAP\_VAR})
constructed from the inverse-variance image produced by MultiDrizzle, and including the
photometric noise from the science data image itself.
In order to flag bad pixels, we made maps denoting blank image areas, 
pixels close to frame boundaries, and the circular detector defect visible
in WFC3/IR images. The maps were referenced using {\tt FLAG\_IMAGE} in SExtractor.
We ran SExtractor with a Gaussian detection filter to identify objects with an area of
at least four connected pixels with a flux level above two times the background rms in
the ACS F475W and F814W images.
The slightly larger value of ${\tt DETECT\_MINAREA}=5$ 
was used for the WFC3/IR F160W image since the subpixel resampling from 
the original pixel scale to 0\farcs1~pix$^{-1}$ during the MultiDrizzle run 
causes more noise correlation between neighboring pixels. 
Separation of blended objects was performed using the SExtractor parameter 
${\tt DEBLEND\_NTHRESH}=32$ and ${\tt DEBLEND\_MINCONT} = 0.005$ and $0.007$ 
for ACS/WFC and WFC3/IR images, respectively. 

The source catalogs extracted from the ACS/WFC F475W and WFC3/IR F160W images
were matched against the ACS/WFC F814W catalog using the source positions to  
remove spurious sources from the multi-band data. 
We estimate total \Iacs\ magnitudes for each object using the \magauto\ values.
For the color estimations, the aperture photometry was performed using apertures
with radii of 3~pixels (0\farcs15 for ACS and 0\farcs30 for 
WFC3/IR data) as in Blakeslee \etal\ (2012). 
Aperture corrections were determined for a typical GC at the Coma distance 
using PSF-convolved King models. The empirical PSFs for ACS/F475W, ACS/F814W, 
and WFC3/F160W bands were produced with the same drizzle parameters, 
including interpolation kernel, pixfrac, and output scale (all of which
have important effects for magnitudes measured within small apertures) as the
science data for each band. 
Our final aperture corrections for the 3-pixel radius SExtractor apertures 
are $-0.24$, $-0.26$, and $-0.28$~mag for \gacs, \Iacs, and \Hwfc, respectively, 
with uncertainties of $0.01$~mag. 

\begin{figure*}
\begin{center}
\includegraphics[width=0.95\linewidth]{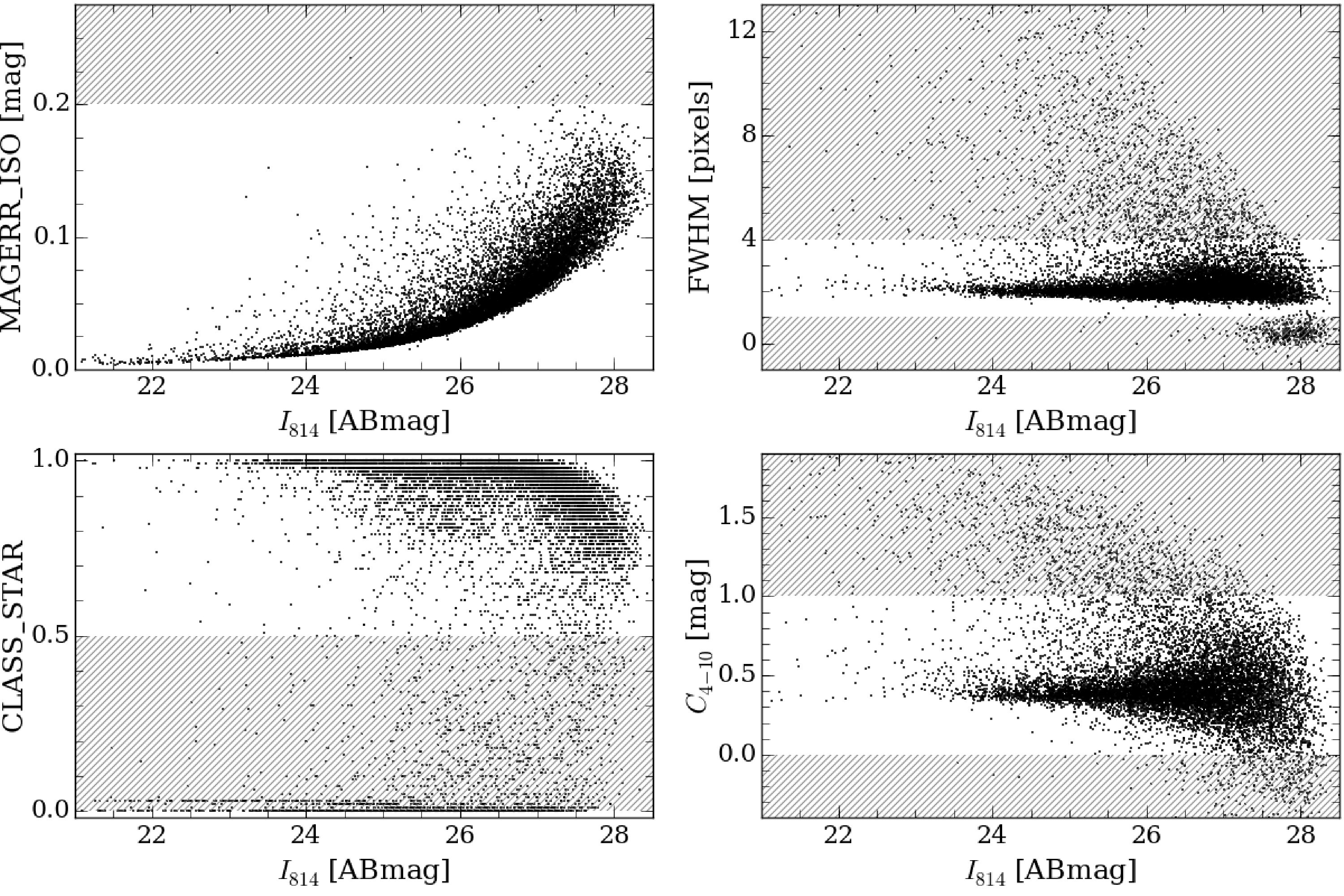}
\caption{
Initial criteria for GC candidate selection as a function of \Iacs\ {\tt MAG\_AUTO}
(an estimate of the total magnitude) from SExtractor. 
Clockwise from top left panel: RMS error for the isophotal magnitude
(scaling inversely with detection signal-to-noise), full width at half-maximum,
magnitude difference between 4- and 10-pixel diameter apertures, and the CLASS\_STAR 
stellarity index values are plotted against {\tt MAG\_AUTO}.
The gray hatched regions mark the parameter ranges over which objects are excluded under
each criterion (see Section~\ref{subsec:detection} for details).
\label{fig:selection}}
\end{center}
\end{figure*}

In order to identify GC candidates, we used the F814W photometric catalog
because of its higher signal-to-noise ratio ($S/N$) than the F475W data,
and larger field of view than the WFC3/IR image.
Prior to classifying candidates as GCs, we required the 
SExtractor parameter ${\tt FLAGS} < 4$ for all three bands in order to exclude 
sources too near the image edges (e.g., Puzia \etal\ 2014). 
To limit our analysis to sources detected with $S/N\,{>}\,5$, 
we require $\magerriso$ (the rms error on the magnitudes within the isophotal
area) in F814W to be less than 0.2~mag.
Figure~\ref{fig:selection} shows our photometric selection criteria for 
probable GCs as a function of the total \Iacs\ magnitudes, for which we
adopt the values of \magauto\ measured with  SExtractor 
(as an additional sanity check, we require the uncertainty in \magauto\
to be less than 1~mag).
As demonstrated in the top left panel of Figure~\ref{fig:selection}, 
the uncertainties on the isophotal magnitudes are smaller than 0.1~mag
for  the majority of the GC candidates
brighter than the turnover of the GC 
luminosity function (GCLF), which is expected to occur at an AB magnitude of
$\Iacs \approx 26.9\pm0.2$~mag at the distance of the Coma cluster
(Peng \etal\ 2009, 2011).  

\begin{figure*}
\begin{center}
\includegraphics[width=1.0\linewidth]{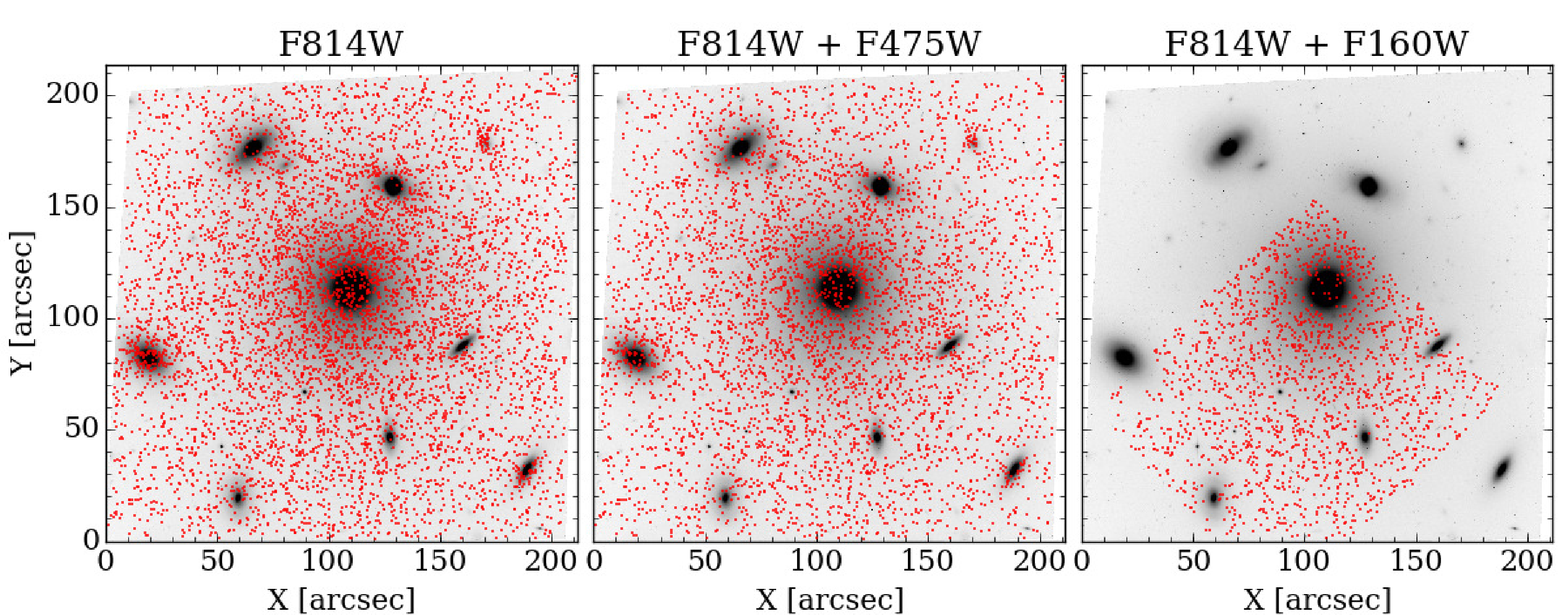}
\caption{
Spatial distributions of GC candidates in the magnitude range of $21.5 < \Iacs <
27.0$~mag (red points) plotted on top of the ACS/WFC F814W image.
The left panel shows positions of the GC candidates selected only from the F814W
photometry.
The middle panel shows GC candidates from the matched F814W and F475W photometric
catalogs with colors $0.5 < \gIacs < 1.6$~mag and color errors $<0.2$~mag.
The right panel shows the positions of the GC candidates from the matched F814W and 
WFC3/F160W photometric catalogs with colors $-0.5 < \IHacs < 1.5$~mag, and color errors
$<0.2$~mag. 
\label{fig:candidates}}
\end{center}
\end{figure*}

The majority of GCs can be treated as point sources in our \hst\ images since 
the mean half-light radius of typical GCs in early-type galaxies, 
$r_{h}\approx3$~pc (Jord\'{a}n \etal\ 2005, 2009; Masters \etal\ 2010), 
corresponds to $\sim\,$0\farcs006 at 100~Mpc.
We therefore required candidate GCs to be compact.
The SExtractor ``stellarity index'' values \classstar\ for all detected objects in
F814W are plotted against the total magnitudes in the bottom left panel of  
Figure~\ref{fig:selection}. The objects with $\classstar > 0.5$ were
classified as point-like sources, and thus possible GCs in Coma.  Since the
\classstar\ parameter is unreliable for the fainter objects, we also adopted
additional criteria, 
based on the measured FWHM and concentration index $C_{4-10}$, to
select faint point-like sources.
The $C_{4-10}$ concentration index was introduced by Peng \etal\ (2011) in
order to select likely GC candidates in Coma; it is defined as the difference
between magnitudes measured in apertures with diameters of 4~pix and 10~pix.
These additional selection criteria are graphically indicated in the right
panels of Figure~\ref{fig:selection}, where it is clear that a large fraction
of the detected sources 
follow tight loci around a \fwhm\ of 2~pix and a $C_{4-10}$ value of 0.4~mag.

We can thus summarize our initial (i.e., from the ACS F814W band,
prior to any color cuts) GC candidate selection criteria as follows: 
$\magerriso{\,<\,}0.2$, 
$1{\,<\,}\fwhm{\,<\,}4$~pix (with 0\farcs05~pix$^{-1}$),
and  $0.0 < C_{4-10} < 1.0$~mag. 
We adopted a relatively broad cut in $C_{4-10}$ in order to include GC
candidates that are more extended than typical GCs.
However, using the above combination of criteria ensures that we select 
robustly characterized compact sources as GC candidates.
In Figure~\ref{fig:candidates} (left panel), we plot the locations in the ACS
F814W image of the 6303 GC candidates selected solely from the F814W photometric
data with magnitudes in the range 
$21.5 < \Iacs < 27.0$~mag. These F814W GC candidates are widely distributed around the 
central cD galaxy NGC\,4874, with localized concentrations around several of the
surrounding cluster galaxies.  

In this work, we also analyze the color properties of the GC candidates,
and for this analysis we impose additional criteria to reject objects that are
likely to be contaminants based on their color.
In matching the F814W-selected candidates with the ACS F475W object catalog,
we imposed a broad color cut of $0.5 < \gIacs < 1.6$~mag 
(e.g., Peng et al.\ 2011) for the GC candidates. We plot 
in the central panel of Figure~\ref{fig:candidates} the 4612 GC candidates
from the left panel that have colors within this range
and color uncertainties less than 0.2~mag.
In matching the F814W candidates to the WFC3/F160W catalog, we restricted 
the colors to $-0.5 < \IHacs < 1.5$~mag (e.g., Blakeslee et al.\ 2012);
again requiring color uncertainties less than 0.2~mag, 
we plot the 1719 GC candidates within this \IHhst\ color interval
in the right panel of Figure~\ref{fig:candidates}. 
Note that the paucity of matched F814W+F160W GC candidates near the galaxy
NGC\,4873 (labeled in Figure~\ref{fig:imgf814w}) occurs because this galaxy is
off the edge of the WFC3/IR field of view (see Figure~\ref{fig:imgf160w})
and was not cleanly subtracted by isophotal modeling; thus, we did not
obtain reliable F160W photometry for objects in its immediate vicinity.

\subsection{Comparison with ACS Coma Cluster Survey}

Photometry in the ACS \gacs\ and \Iacs\ bands for GCs in the region around NGC\,4874 was
previously published by Peng \etal\ (2011) using data from \HST\ program GO-10861, ACSCCS.
Our exposure time in F814W is 7.4~times longer than that obtained by the ACSCCS, implying 
a $S/N$ about 2.7~times greater, or a limiting magnitude more than 1~mag deeper in this band.
For F475W, because we incorporated the ACSCCS exposures into our stacked image, our
exposure time is nearly a factor of two longer ($\sim40$\% higher $S/N$) 
than for the ACSCCS data alone (the images did not overlap completely
because they were taken at different orientations).
Since the addition of the ACSCCS F814W data would
have increased our $S/N$ by $\lta7\%$, we opted not to include those data in
our stacked image in that bandpass. Peng \etal\ (2011) performed source photometry on the
galaxy-subtracted ACSCCS images using SExtractor with 3~pixel radius apertures and then
selected GC candidates based on color and source concentration; thus, the analysis was
quite similar to our own and can be used as a straightforward check on our photometry.

\begin{figure}
\begin{center}
\includegraphics[width=0.96\linewidth, trim=0 0 0 -0.5cm]{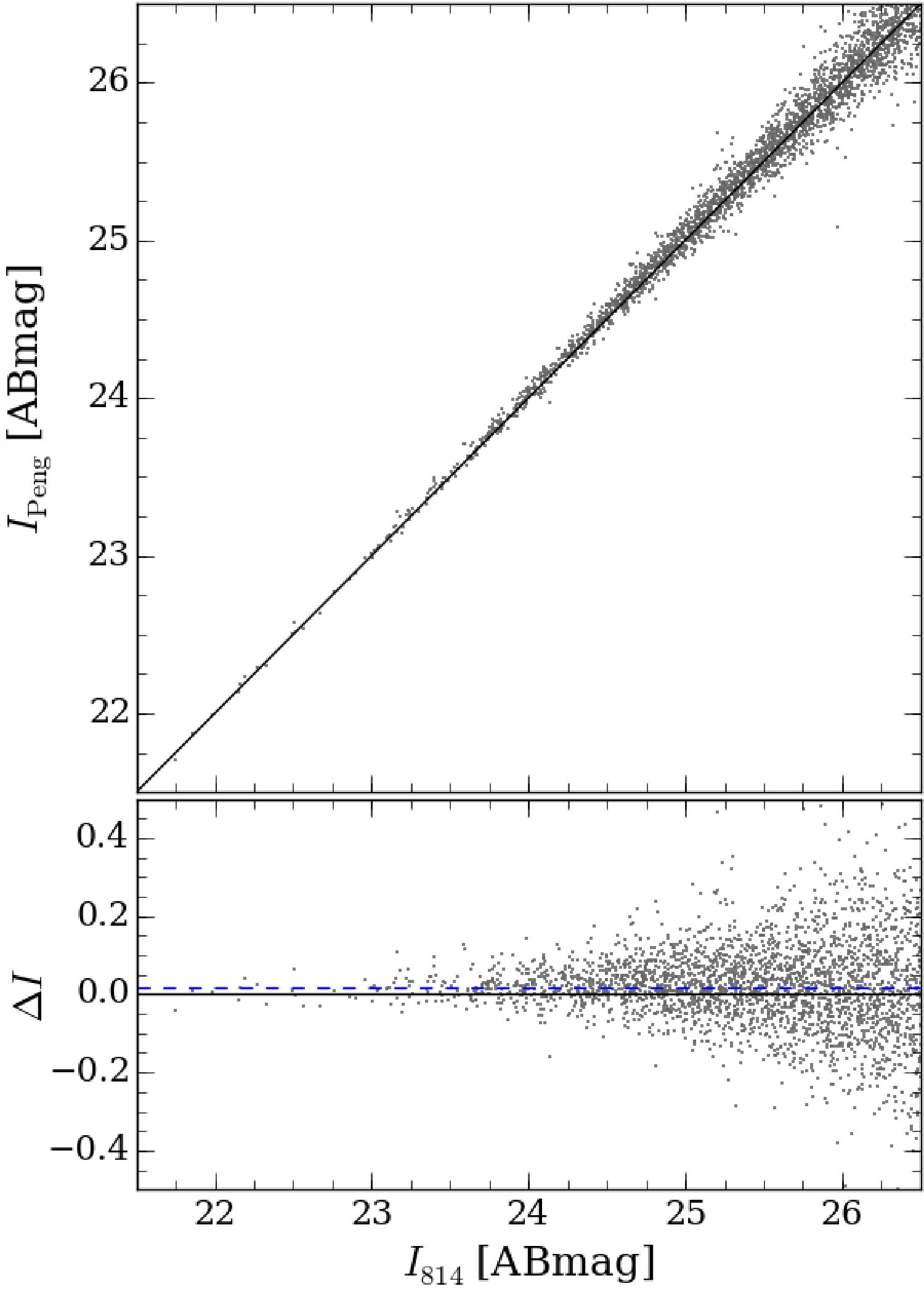}
\caption{
Comparison of 3-pixel radius aperture magnitudes in the ACS F814W passband from 
the photometry of Peng \etal\ (2011), denoted by \Ipeng, with those from our 
photometry. The points are objects in common for the two data sets. 
The black solid line in the upper panel represents equality, 
and is not a fit. In the lower panel, 
the magnitude differences $\Delta{I}=\Ipeng-\Iacs$ are plotted as a function of 
our \Iacs\ magnitude. The black solid line shows a zero magnitude difference, 
while the blue dashed line indicates the median offset in $\Delta{I}$ of 0.017~mag. 
\label{fig:acsccsImag}}
\end{center}
\end{figure}

Figure~\ref{fig:acsccsImag} shows a comparison of \Iacs\ magnitudes from Peng \etal\ 
(2011) with those of the present study; for consistency, we compare the magnitudes
without correction for Galactic extinction. The data for this comparison (unlike the
case for \gacs) are fully independent. The top panel of the figure shows that the
overall agreement is very good over a range of 5~mag;
the slope of the residuals over this magnitude range is consistent with zero.
Peng \etal\ cut their GC
selection at $\Iacs=26.5$~mag, in part because the photometric error in their $C_{4-10}$
concentration parameter became too large to distinguish point sources and background
galaxies at about this magnitude; as expected from the increased depth, our F814W data
are able to distinguish point sources from extended objects to about 1~mag fainter
(compare their Figure~2 with our Figure~\ref{fig:selection}).

\begin{figure*}
\begin{center}
\includegraphics[width=0.96\linewidth]{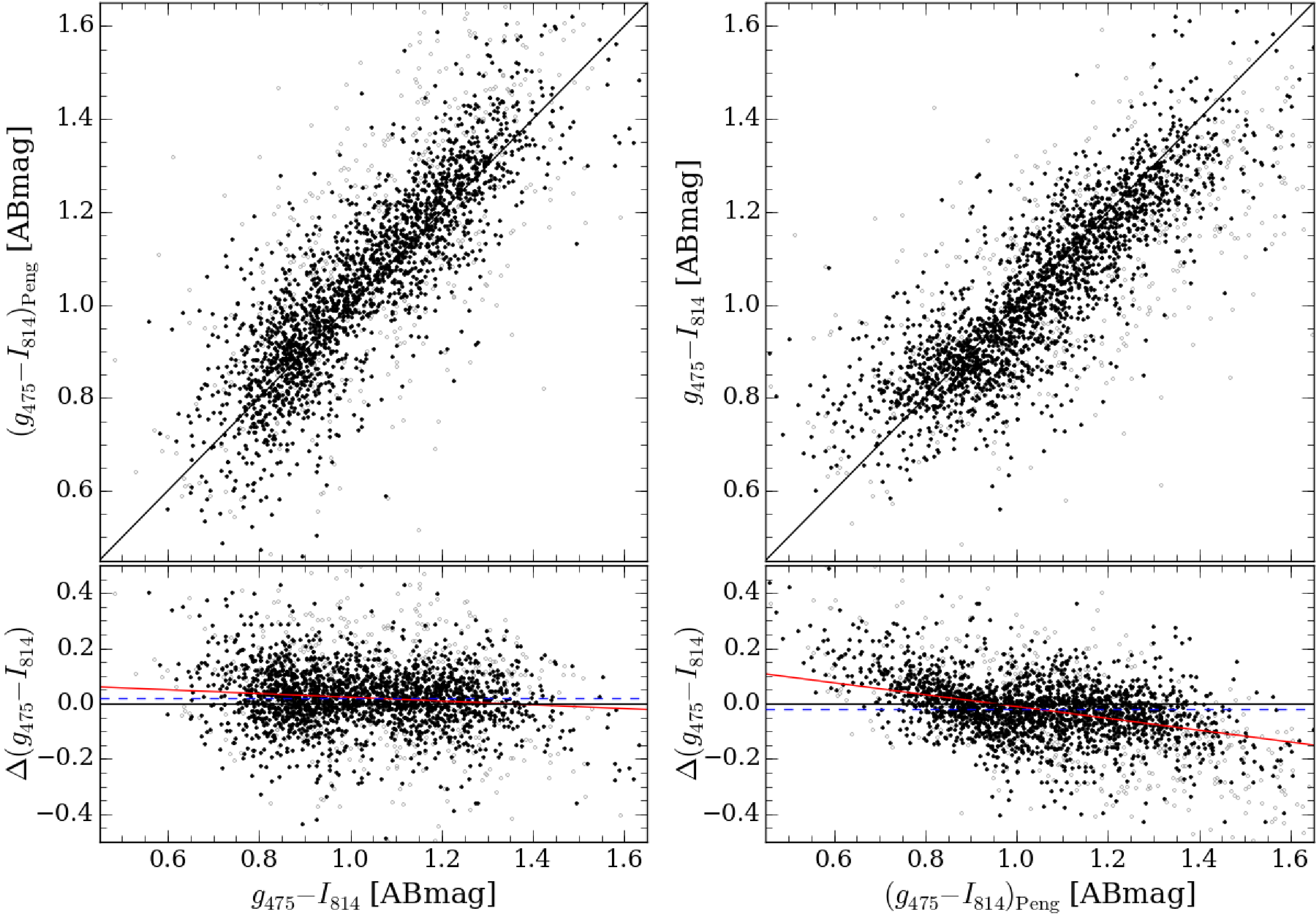}
\caption{
Comparison between our \gIacs\ color values and those from the photometry of 
Peng \etal\ (2011), denoted by \gIpeng. Black points have color errors 
in the Peng \etal\ photometry smaller than 0.2~mag. The objects with larger 
color uncertainties are marked by gray points. The black solid lines in the upper 
panels represent the one-to-one relation. The lower panels show the differences 
in the colors between these two data sets. In each lower panel, the zero 
and median difference values are marked by the black solid line and the blue 
dashed line, respectively. The red solid lines show the robust linear relations 
given in the text.
\label{fig:acsccsgIcolor}}
\end{center}
\end{figure*}

The lower panel of Figure~\ref{fig:acsccsImag} shows the residuals
$\Delta{I}=\Ipeng{\,-\,}\Iacs$ for 2673 point sources (selected based on our
measurements) in common between the two data sets down to $\Iacs=26.5$~mag (again, from
our measurement). The mean offset in \Iacs\ is $0.014\pm0.002$~mag, with a scatter of 0.105~mag,
and the median offset is 0.017~mag, in the sense that the ACSCCS magnitudes are slightly
fainter. If we limit the range to $21.5<\Iacs<25.5$~mag, then the number of sources is reduced by 
approximately half to 1308, with both a mean and median offset of 0.022~mag, and a scatter of 0.063~mag.
Peng \etal\ (2011) calibrated their photometry using the AB zero points from
Sirianni \etal\ (2005); however, adopting the calibration from the online ACS zero
point calculator for the appropriate date of the observations would decrease the size of
the offset by only 0.002~mag. Given the uncertainty in the time-dependence of the zero
point (Bohlin 2012), the lack of CTE correction for the ACSCCS data, the difference in the drizzle
parameter settings (e.g., linear versus lanczos3 interpolation kernels), the possibility
of small focus variations (e.g., Jee \etal\ 2007), and the much
greater depth of our F814W observations (which could result in subtle differences in the
SExtractor photometry), we consider the systematic offset of $\lta0.02$~mag to be
reasonable. The scatter in the residuals (dominated by the much shallower ACSCCS
measurements) increases as expected at fainter magnitudes, but there is no evidence for
a systematic trend in the residuals with magnitude.

Figure~\ref{fig:acsccsgIcolor} compares the \gIacs\ colors for point sources in common
between Peng \etal\ (2011) and the present study over a range in \Iacs\ from 21.5 to
26.5~mag. The left and right panels show, respectively, the comparison as a
function of our and the ACSCCS color measurements. In both panels, the black points
represent sources in Peng \etal\ (2011) with estimated color errors $<0.2$~mag, while the
gray points show sources with color errors larger than 0.2~mag. Considering all the
points, black and gray, over the plotted color range of $0.45<\gIacs<1.65$~mag, 
the median offset is 0.026~mag, the rms scatter is 0.13~mag, and the biweight scatter 
(more robust against outliers) is 0.12~mag.
Considering just the black points, the median offset is 0.020~mag, the rms scatter is 0.11~mag,
and the biweight scatter is 0.10~mag.
The sense of the offset is that the ACSCCS \gIacs\ colors are slightly redder than ours;
if we were to recalibrate the Peng \etal\ photometry
using the online ACS Zeropoints Calculator, the ACSCCS colors would become bluer by 0.021~mag,
reducing the median color offset for the black points to 0.001~mag. However, because of
observational error, the observed offset also has a dependence on color.

The lower panels of Figure~\ref{fig:acsccsgIcolor} show the color differences
$\Delta{(\gIacs)}$ (defined as $y$-axis color minus $x$-axis color) plotted as a
function of both our colors and the ACSCCS colors, which we label $(\gIacs)_{\rm Peng}$.
The solid red lines show robust linear regressions for the black points 
in these panels; the slopes of the $\Delta{(\gIacs)}$ regression lines
are $-0.068\pm0.012$ and $-0.214\pm0.010$ when fitted versus our \gIacs\ colors and versus
$(\gIacs)_{\rm Peng}$, respectively.  
Thus, the slope is more than a factor of three steeper when fitted as a
function of the ACSCCS colors.  This is understandable in light of the
larger measurement errors for those colors. 
Since the vast majority of these objects are GCs, which intrinsically define a fairly
narrow color range $0.7\lta\gIacs\lta1.4$~mag, the scattering of the colors
outside this color range primarily results from photometric errors, which are
larger for the shallower ACSCCS measurements; thus, this error-induced slope is
larger when plotted as a function of the ACSCCS colors. 
We find that we can reproduce the slopes and scatters
in Figure~\ref{fig:acsccsgIcolor} if we assume Gaussian errors with $\sigma=0.055$~mag
for our color measurements and $\sigma=0.107$~mag the ACSCCS colors. 
For comparison, the median estimated color errors in the two catalogs are
0.063~mag and 0.098~mag, respectively. This suggests that our quoted errors may
be slightly overestimated and the ACSCCS color errors slightly underestimated,
but only by about 10\% in each case.

We conclude that our measurements agree well with the ACSCCS photometry from Peng \etal\ (2011). 
The much greater exposure time of our \Iacs\ imaging allows us to reach about
1~mag fainter in this bandpass, while our \gIacs\ color errors for GC candidates are
approximately a factor of two smaller than for the ACSCCS data. Systematic offsets in
photometry are $\lta0.02$~mag. In addition, our program adds deep \Hwfc\ photometry over the
area of WFC3/IR field, which was not available for the earlier study.

\section{Discussion}
\label{sec:discussion}
\subsection{Color-Magnitude Diagrams and ``Tilts''}
\label{sec:cmds}

As discussed in the Introduction, GC systems of massive galaxies generally follow
bimodal distributions in optical color. However, the peaks in the color distribution can
vary with the magnitude range of the GCs considered. For instance, Ostrov \etal\ (1998)
and Dirsch \etal\ (2003) found that for GCs more than 2~mag brighter than the turnover
of the GCLF in the galaxy NGC\,1399, the blue and red peaks merged together into a
single broad distribution. More generally, the mean color of the blue GCs tends to get
redder at brighter magnitudes (Harris \etal\ 2006; Mieske \etal\ 2006, 2010; Strader
\etal\ 2006; Harris 2009), 
possibly indicating an increasing mean metallicity with GC
luminosity. This effect, known alternately as the GC color-magnitude relation,
mass-metallicity relation, or informally as ``the blue tilt'' (a ``tilt'' of the blue
peak towards a redder mean color at bright magnitudes) is most generally believed to be 
a consequence of
self-enrichment within the most massive GCs (e.g., Bailin \& Harris 2009).

\begin{figure*}
\begin{center}
\includegraphics[width=0.85\linewidth]{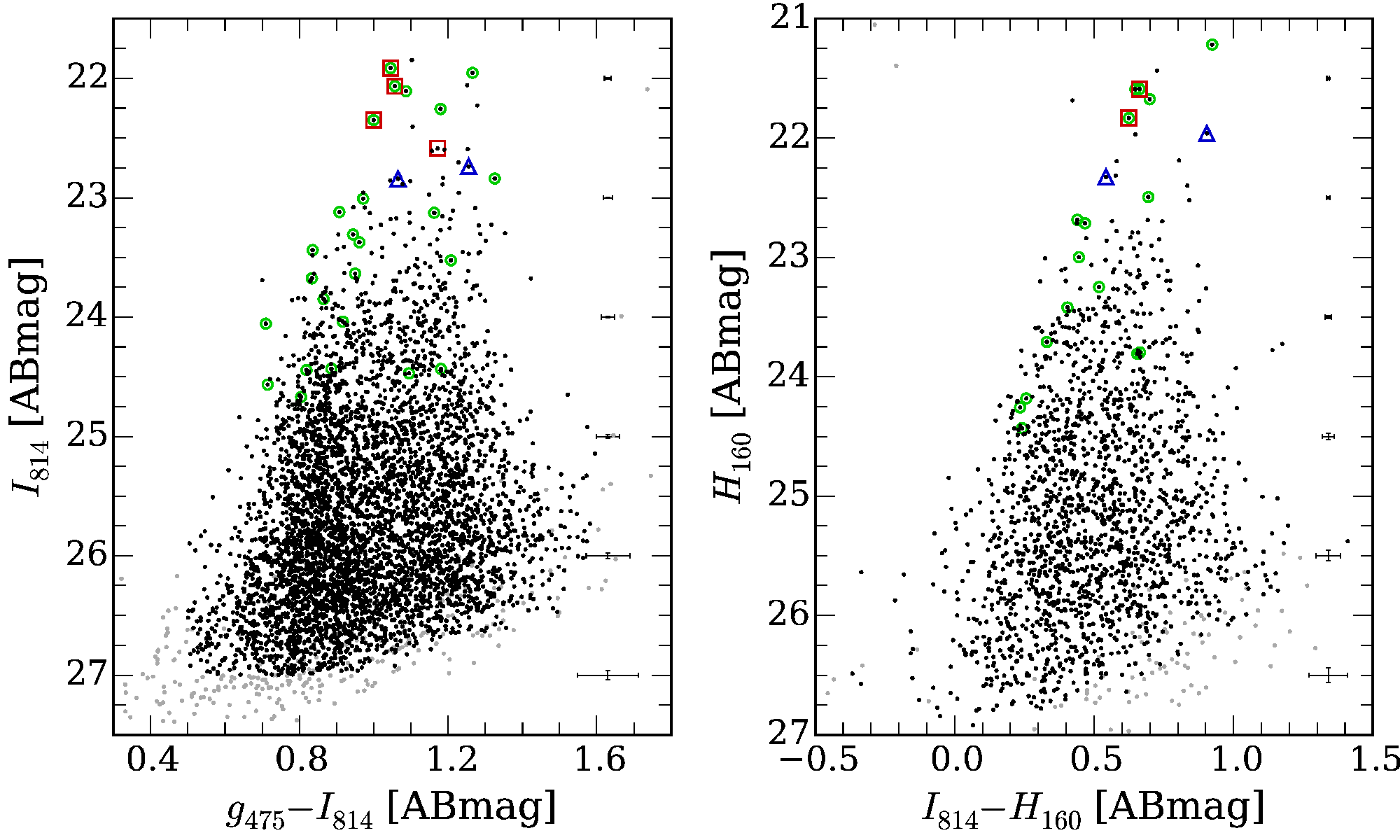}
\caption{
The optical color-magnitude diagram for GC candidates in the NGC\,4874 field 
from ACS/WFC F475W and F814W imaging data (left) and the optical-NIR 
color-magnitude diagram for GC candidates from ACS/WFC F814W and WFC3/IR F160W 
imaging data (right). The final GC samples from Figure~\ref{fig:candidates} 
are plotted in black; the gray points were excluded from further analysis. 
Error bars (near the right edge of each panel)
represent the mean errors of the magnitudes and colors in a magnitude bin. 
Red squares show ultra-compact dwarf galaxies (UCDs, classified purely on the basis of
luminosity and color) spectroscopically confirmed 
as members of the Coma cluster, while blue triangles indicate likely UCDs with uncertain redshifts
(Chiboucas \etal\ 2011). Green circles mark photometrically classified UCDs or
``dwarf-globular transition objects'' from Madrid \etal\ (2010). These objects all have
photometric properties consistent with being extensions of the GC population, and
we do not attempt to exclude them from our analysis. 
\label{fig:cmd}}
\end{center}
\end{figure*}

Figure~\ref{fig:cmd} displays the color-magnitude diagrams (CMDs) for GC candidates in
NGC\,4874 in both the optical \gIacs\ (left panel) and optical-NIR \IHhst\ (right panel) colors.
We have marked in these panels the objects, included in our GC selection, that are
spectroscopically confirmed (red squares) or possible (blue triangles)
ultra-compact dwarfs (UCDs) from the study of Chiboucas \etal\ (2011).
In this case, UCDs are defined simply as compact stellar systems with colors
similar to GCs and absolute $R$ magnitude $M_R{\,<\,}-11$~mag.
Because of the larger ACS/WFC field, there are six of these objects in the optical CMD
of the left panel, but four in the right panel (in both cases, two of the UCDs are
uncertain Coma members based on their spectra). We also indicate objects 
(green circles) that were selected
by Madrid \etal\ (2010) based on ACSCCS imaging as candidate (lacking spectroscopic
confirmation) UCDs or ``dwarf-globular transition objects,'' defined as objects
having GC-like colors and half-light radii in the range of 10~to 100~pc (if located at
the distance of the Coma cluster).
Although there are more objects in the left panel, and the ridge-line of the blue GC
component is also much more distinct in the \gIacs\ color, overall the CMDs appear fairly
similar over a range of 5~mag in luminosity, with an overall tilt towards redder colors
at the brightest magnitudes where objects tend to be classified as UCDs.

\begin{figure*}
\begin{center}
\includegraphics[width=0.84\linewidth]{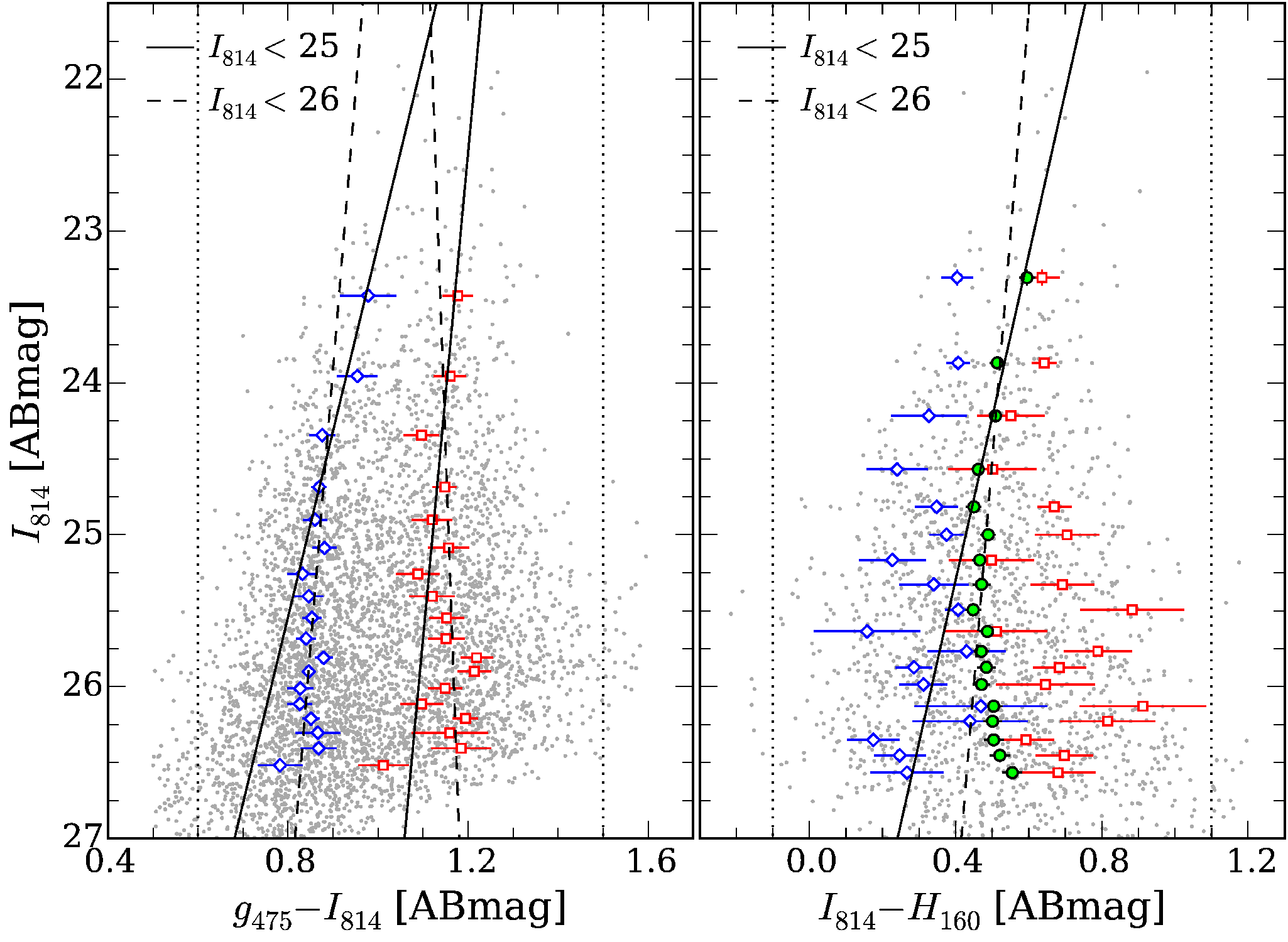}
\caption{
\Iacs\ vs. \gIacs\ color-magnitude diagram (left) and \Iacs\ vs. \IHacs\ 
color-magnitude diagram (right). The black points in Figure~\ref{fig:cmd} are 
plotted in gray in these diagrams for clarity. We subdivided each CMD into 
eighteen magnitude bins, each with a fixed number of data points, except the
brightest two bins in each panel, for which the number is half that of the other bins.
In the left panel, the positions of the first and second peaks (mean
positions) in the GMM double Gaussian model for each magnitude bin are plotted with blue
open diamonds and red open squares, respectively.  
The error bars represent the uncertainties on each peak calculated from
the non-parametric bootstrap resampling by the GMM algorithm.
In the right panel, the green circles with error bars 
indicate the positions of the average peaks for each magnitude bin. We also plot 
the positions of the double peaks from the GMM fits, which are marked 
by blue open diamonds and red open squares, 
along with the error bars from the GMM bootstrap resampling.
The thick solid and dashed lines indicate the linear fits to the peak positions 
when the faintest magnitudes used in the each fit are $\MIacs=-10.0$~mag and 
$-9.0$~mag, respectively. For the fits, we adopted the blue and red color limits 
marked by the vertical dotted lines. 
\label{fig:cmdtilt}}
\end{center}
\end{figure*}

In order to quantify the degree of ``blue tilt'' in NGC\,4874, we binned the GC
candidates by magnitude and applied the Gaussian Mixture Modeling (GMM) code of Muratov
\& Gnedin (2010) to each bin.
Figure~\ref{fig:cmdtilt} shows CMDs similar to the previous figure, but now using \Iacs\
for the magnitude in both cases, and showing the locations of the color peaks from the
bimodal GMM decompositions for eighteen bins in magnitude down to $\Iacs\approx26.5$~mag.
Because the GMM algorithm can be sensitive to objects that are
scattered into the tails of the distribution by observational errors (which
increase at fainter magnitudes), we have restricted these magnitude-grouped
samples to the color ranges $0.6<\gIacs<1.5$~mag 
and $-0.1<\IHhst<1.1$~mag, indicated by the dotted lines in Figure~\ref{fig:cmdtilt}.
For the \Iacs\ versus \gIacs\ CMD (left panel), each bin has 240 GCs, while 
for the \IHacs\ CMD (right panel), each bin has 90 GCs; the exceptions are the brightest
two bins in each panel, which have only half the number of GCs as the other bins.
In the left panel of Figure~\ref{fig:cmdtilt}, there is clear evidence for a ``blue tilt,''
as well as some suggestion of a ``red tilt'' for the peak positions at magnitudes
$\Iacs<25$~mag, corresponding to absolute $\MIacs<-10$~mag. Linear fits to the red and
blue peak positions for bins brighter than this magnitude are shown by the solid black
lines, defined by the following relations:
\begin{eqnarray}
(\gIacs)_{\rm blue} &=& (0.88\pm0.01) \,-\, (0.082\pm0.020) \nonumber \\
~&~& {\times}(\Iacs-24.5) , \; \MIacs < -10.0\,; \label{eq:BgIvsIbright}\\
(\gIacs)_{\rm red} &=& (1.14\pm0.01) \,-\, (0.031\pm0.023) \nonumber \\
~&~& {\times}(\Iacs-24.5) , \; \MIacs <-10.0\,. 
\end{eqnarray}
The error bars here reflect the statistical uncertainties in the parameters 
from the linear fits.

The slope $\gamma_{814,{\rm blue}}\equiv{d(\gIacs)}/{d\Iacs}=-0.082$ of the blue tilt is 
highly significant (4-sigma) and is among
the steepest observed to date. For comparison, Mieske \etal\ (2006, 2010) found
$\gamma_{850,{\rm blue}}\equiv{d(\gzacs)}/{d\zacs}=-0.042\pm0.015$ for M87 in Virgo and 
$\gamma_{850,{\rm blue}}=-0.088\pm0.025$ for NGC\,1399 in Fornax, 
the central giant ellipticals in each cluster. Using the observed
relationship for GCs in NGC\,1399, $\gIacs = 0.13 + 0.75(\gzacs)$ from 
Blakeslee et al.\ (2012), this would imply 
$\gamma_{814,{\rm blue}}=-0.032$ and $-0.065$ for M87 and NGC\,1399, respectively.
Further, NGC\,4472 (M49), the brightest galaxy in Virgo, has no significant blue tilt at
all; thus, the color-magnitude trend in NGC\,4874 is exceptionally steep
compared to the Virgo and Fornax clusters. 
This may be related to an abundance of UCDs in the dense core of the Coma cluster.
It is clear that the tilt becomes greater for objects
at $\Iacs<23.5$~mag ($M_{814}<-11.5$~mag), where the sample may be dominated by UCDs,
which tend to have colors intermediate between the blue and red peaks of the
optical GC color distribution (e.g., Liu et al.\ 2015).  It is likely that UCDs 
represent a mix of stripped galactic nuclei and luminous GCs;
as already indicated in Figure~\ref{fig:cmd}, we have not attempted to exclude
UCDs from our sample if they satisfy our selection criteria.

The derived color-magnitude slope becomes markedly less steep when the fit is
extended to fainter GCs. The dashed lines in Figure~\ref{fig:cmdtilt}
indicate the following linear fits to the peaks with $\Iacs<26$~mag ($M_{814}<-9$~mag):
\begin{eqnarray}
(\gIacs)_{\rm blue} &=& (0.88\pm0.01) \,-\, (0.027\pm0.010) \nonumber \\
~&~&{\times}(\Iacs-24.5) , \; \MIacs < -9.0\,; \label{eq:BgIvsIfaint}\\
(\gIacs)_{\rm red} &=& (1.15\pm0.01) \,+\, (0.012\pm0.015) \nonumber \\
~&~&{\times}(\Iacs-24.5) , \; \MIacs <-9.0\,.
\end{eqnarray}
Thus, when the fit is extended by one magnitude, the slope of the
color-magnitude trend for blue peak positions is reduced by a factor of three. 
The shallower slope over the wider magnitude range reflects the nonlinearity
of the color-magnitude tilt (e.g., Harris 2009; Mieske \etal\ 2010), 
which may result from a minimum mass threshold for self-enrichment.
For the red GC peak, the fitted slope over this broader magnitude range is essentially zero.

We can estimate the scaling of metallicity $Z$ with the GC luminosity $L$ in the
\Iacs\ bandpass using the empirical broken-linear calibration from Peng \etal\ (2006)
for the metallicity as a function of \gzacs\ color. Since the ``tilt'' occurs for the
blue GC population, we use the linear relation appropriate for the blue GCs:
$\feh = -6.21 - (5.14\pm0.67)(\gzacs)$. This is the same relation used by Mieske \etal\ (2010) 
for deriving the mass-metallicity scaling from their blue tilt measurements in
the ACS Virgo and Fornax Cluster Survey data
(C{\^o}t{\'e} et al.\ 2004; Jord{\'a}n et al.\ 2007). 
Coupled with the above relation between \gIacs\ and \gzacs, and our measurement of 
$\gamma_{814,{\rm blue}}=-0.082\pm0.020$ for $\MIacs<-10$~mag, we find
$Z\propto L^{1.4\pm0.4}$ at these highest luminosities, or
if we assume a constant mass-to-light ratio for blue-peak GCs as in Mieske \etal,
then $Z\propto M_{\rm GC}^{1.4\pm0.4}$ for the scaling with GC~mass.
Of course, if we use the slope
$\gamma_{814,{\rm blue}}=-0.027\pm0.010$ 
from the linear fit extending to $\MIacs=-9$~mag, then
the mean mass-metallicity scaling over this magnitude range becomes 
$Z\propto M_{\rm GC}^{0.5\pm0.2}$, again reflecting the nonlinearity of the relation.

For the \Iacs\ versus \IHhst\ CMD
(Figure~\ref{fig:cmdtilt}, right panel), we find no significant evidence for a ``tilt'' in the colors
of either the red or blue peaks from the GMM bimodal decompositions within the magnitude
bins. This may be because of the poorer statistics and/or weaker separation of blue and
red GCs for this optical-NIR color. Notably, however, we do find a significant trend for
the overall mean GC color (based on the unimodal GMM fit) to become redder for the
brighter magnitude bins. The solid line in this panel is a fit to the \IHhst\ unimodal peak
positions for bins with $M_{814}<-10$~mag; the dashed line again extends the fit one
magnitude fainter than this and is significantly less steep.
The fits are given by the following relations:
\begin{eqnarray}
(\IHhst)_{\rm mean} &=& (0.47\pm0.01) \,-\, \,(0.093\pm0.013) \nonumber \\
~&~&{\times}(\Iacs-24.5) , \; \MIacs <-10.0\,; \label{eq:IHvsIbright}\\
(\IHhst)_{\rm mean} &=& (0.50\pm0.01) \,-\, \,(0.034\pm0.010) \nonumber \\
~&~&{\times}(\Iacs-24.5) , \; \MIacs <-9.0\,.\label{eq:IHvsIfaint}
\end{eqnarray}
The slope of this ``mean tilt'' in \IHhst\ for $\MIacs < -10$, $\Iacs<-25$~mag, is highly significant.
The color sequence appears nearly vertical at magnitudes fainter than this, although the slope 
of the fit for $\MIacs<-9$~mag (dashed line) remains significant because of the brightest bins with
their increasingly steep slope.
Although the relation between \IHhst\ and metallicity has not been empirically
calibrated for extragalactic GC systems, we can check for consistency by using 
the linear version of the relation between \IHhst\ and \gIacs\ derived in 
Sec.~\ref{subsec:ccr} below, 
which has a slope $d(\IHhst)/d(\gIacs) = 1.13\pm0.04$.
Combining this with the same set of relations between \gIacs, \gzacs, and \feh\ as above
(although the adopted \feh\ transformation is only strictly applicable for blue GCs),
we can derive the mass-metallicity scaling from the fitted slopes
${d(\IHhst)}/{d\Iacs}$ in Eqs.\,(\ref{eq:IHvsIbright}) and~(\ref{eq:IHvsIfaint}).
For $\MIacs<-10$~mag, the result is again $Z\propto M_{\rm GC}^{1.4}$, and the
exponent again drops to $\sim0.5$ if we use the \IHhst\ fit extending to $\MIacs=-9$~mag.

\begin{figure}
\begin{center}
\includegraphics[width=1.0\linewidth]{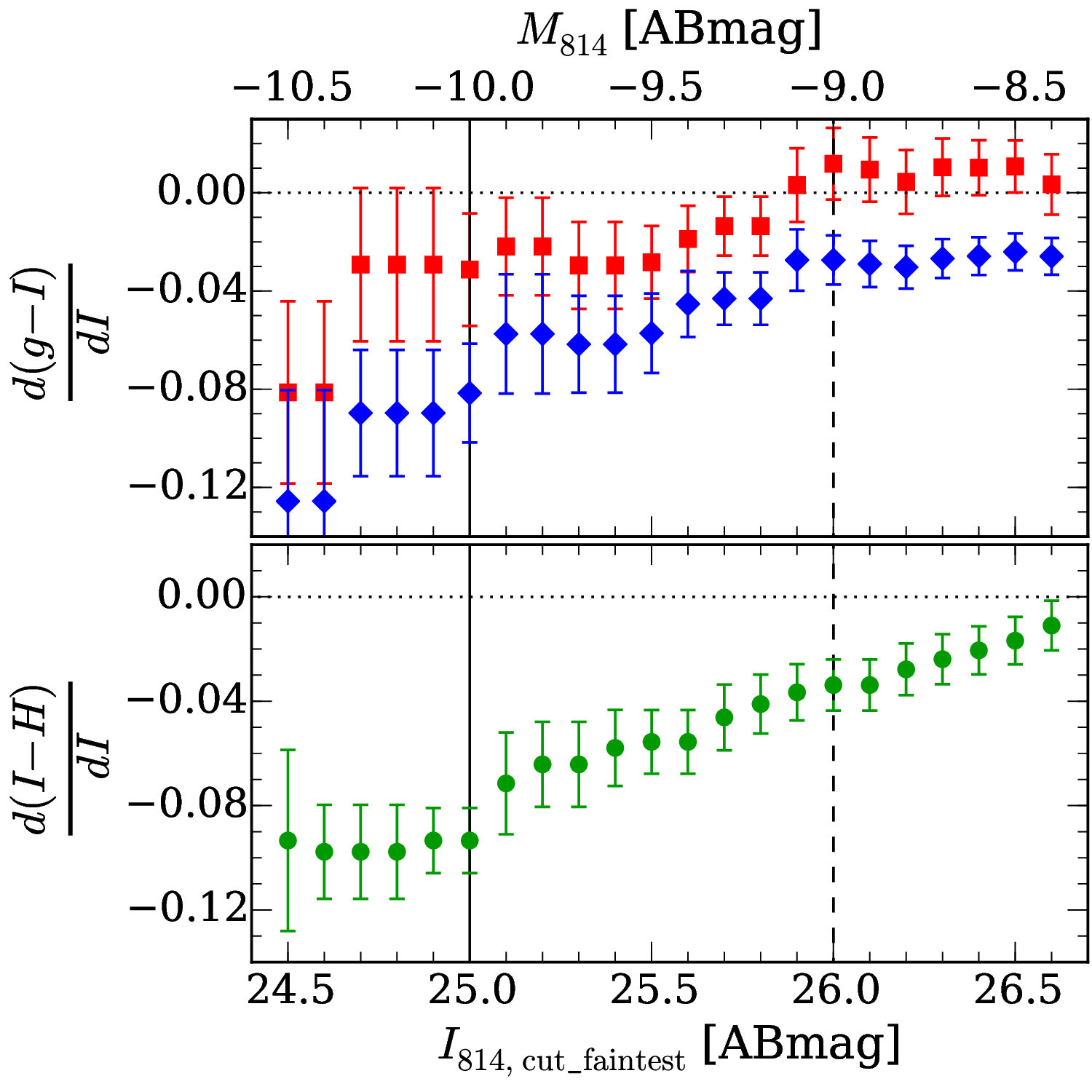}
\caption{
Fitted slopes as a function of the faintest magnitude cut in the ACS F814W passband. 
The red squares and blue diamonds in the top panel are the slope 
values of the linear fits to the red and blue peak positions, respectively,
in the \gIacs\ versus \Iacs\ CMD shown in Figure~\ref{fig:cmdtilt} 
for varying limiting magnitudes of the fit.
The green points in the bottom panel are the slopes of the linear fits 
to the overall mean positions in the \IHhst\ versus \Iacs\ CMD in
Figure~\ref{fig:cmdtilt} for varying limiting magnitudes of the fit.
The vertical solid and dashed lines indicate the limiting magnitude values for the
corresponding linear fits shown in Figure~\ref{fig:cmdtilt}. 
The trends seen here towards shallower slopes with fainter limiting magnitudes indicate that
the relations between mean color and magnitude are nonlinear, with the color peak positions
in the CMDs becoming more vertical at fainter magnitudes; put another
way, the ``tilts'' are only significant for the brightest GCs.
\label{fig:tiltslope}}
\end{center}
\end{figure}

The equality in the exponents of the mass-metallicity relations derived from \gIacs\
and \IHhst\ may seem strange, given that in former case it is based on the trend in
the blue GC component with magnitude, while for the latter it is based on the overall
mean \IHhst\ trend with magnitude. In fact, it is somewhat fortuitous.
The ratio of the slope for the mean \IHhst\ in Eq.~(\ref{eq:IHvsIbright}) 
to the slope for the blue peak in Eq.~(\ref{eq:BgIvsIbright}) is $1.13\pm0.28$;
the corresponding ratio for Eqs.~(\ref{eq:IHvsIfaint}) and~(\ref{eq:BgIvsIfaint}) is
$1.26\pm0.42$.  Both of these are in statistical agreement with the color-color slope
found below in Sec.~\ref{subsec:ccr}.
This agreement can be understood, at least in part, from the fact that at
progressively brighter magnitudes, the proportion of red-peak to blue-peak GCs
increases in the \gIacs\ histogram, as shown in Sec.~\ref{sec:distribs} below. 
Thus, the overall mean slope of \gIacs\ versus \Iacs\ will be steeper than the average of
the red and blue slopes.  For completeness, we also fitted the overall mean
\gIacs\ color-magnitude relations, finding:
\begin{eqnarray}
(\gIacs)_{\rm mean} &=& (1.02\pm0.01) \,-\, \,(0.060\pm0.014) \nonumber \\
~&~&{\times}(\Iacs-24.5) , \; \MIacs <-10.0\,; \label{eq:gImeanbright}\\
(\gIacs)_{\rm mean} &=& (1.03\pm0.01) \,-\, \,(0.027\pm0.008) \nonumber \\
~&~&{\times}(\Iacs-24.5) , \; \MIacs <-9.0\,.\label{eq:gImeanfaint}
\end{eqnarray}
In both cases, the slope is steeper than the average of the red and blue slopes
derived for the equivalent magnitude limits. In fact, the slope we find for the mean
trend in Eq.~(\ref{eq:gImeanfaint}) is the same as that for the blue tilt in
Eq.~(\ref{eq:BgIvsIfaint}). However, the conversion of these mean trends to a
mass-metallicity scaling relation is less straightforward because there is a
change in the slope of the color-metallicity relation at intermediate
colors (e.g., Peng et al. 2006; Usher et al. 2012). 

Figure~\ref{fig:tiltslope} explores in more detail the dependence of the slope of the
color-magnitude tilts as a function of the faint limit of the linear fits. The slope of
the blue peak in \gIacs\ remains significant regardless of the magnitude limit, while the
slope for the red peak appears significant at the $2\sigma$ level only when the brightest
two or three bins are considered, those with $M_{814}<-10.3$~mag. For \IHhst\ (lower panel),
although the separation into blue and red peaks is weak (as quantified in the
following section), for bins with $M_{814}<-8.8$~mag, the slope of the overall trend
towards a redder mean color (and thus metallicity) at brighter magnitudes is quite
significant. However, the magnitude of the slope decreases continuously from
$M_{814}\approx-10$ to $M_{814}\approx-8.5$~mag, again illustrating the nonlinearity of
the trend.

\subsection{Color Distributions}
\label{sec:distribs}

\begin{figure*}
\begin{center}
\includegraphics[width=13.19cm]{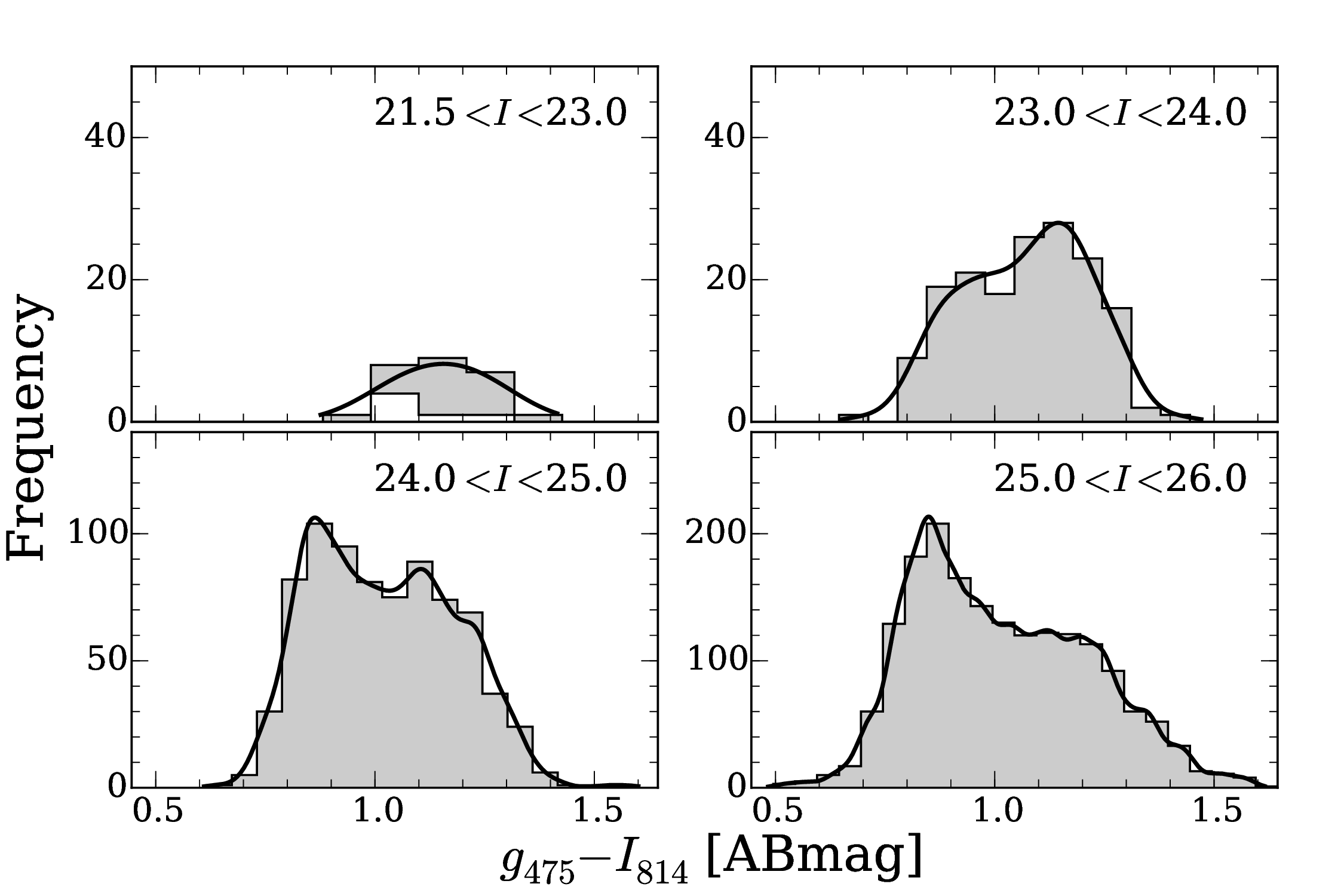}
\caption{
Histograms of \gIacs\ colors for GC candidates in the left panels of 
Figures~\ref{fig:cmd} and~\ref{fig:cmdtilt} over different {\it I}-band magnitude ranges. 
The smooth Gaussian kernel density estimates are overplotted by thick solid curves. 
The hatched histogram is for UCDs (both confirmed cluster members and uncertain 
ones) from Chiboucas \etal\ (2011).
\label{fig:hist1}}
\end{center}
\end{figure*}

\begin{figure*}
\begin{center}
\includegraphics[width=13.19cm]{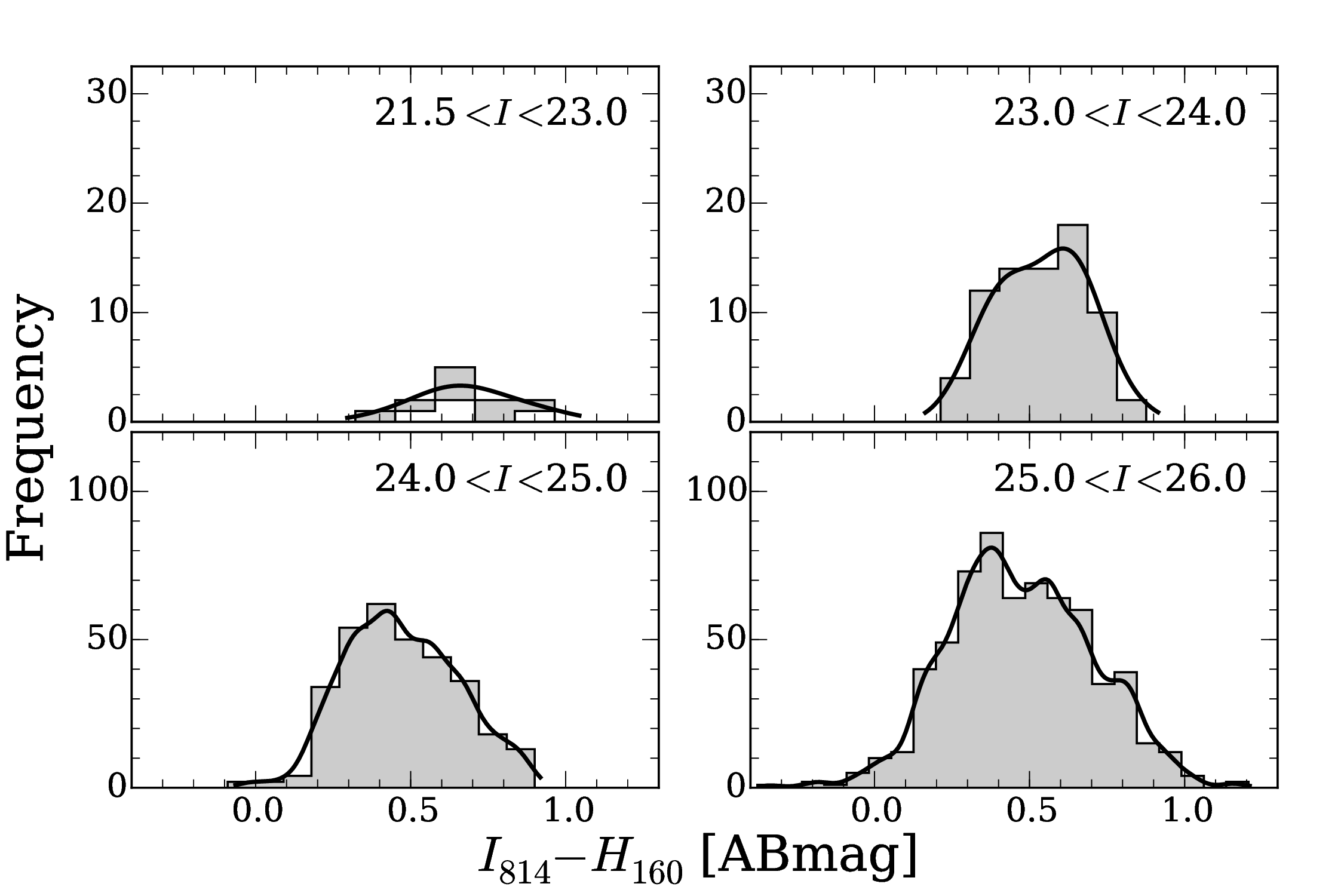}
\caption{
Same as Figure~\ref{fig:hist1} but for \IHacs\ colors for GC candidates 
in the right panels of Figures~\ref{fig:cmd} and~\ref{fig:cmdtilt}. 
\label{fig:hist2}}
\end{center}
\end{figure*}

As discussed in the previous section, the NGC\,4874 GC candidates exhibit a distinct
color-magnitude relation, at least at $\Iacs<25$~mag. Consequently, their color distributions 
should vary as a function of luminosity. 
Figure~\ref{fig:hist1} plots the \gIacs\ optical
color histograms for the GC candidates in four different bins in \Iacs\ magnitude; the
distributions differ markedly from each other. In the brightest bin, consisting
of objects at least 4~mag brighter than the expected turnover of the GCLF, the
distribution is relatively red and broad, with no evidence for bimodality. The
spectroscopic sample of UCDs from Chiboucas \etal\ (2011) is weighted towards the blue
side of this distribution, but the sample is small and incomplete (see
Figure~\ref{fig:cmd}), and selection effects could play a role. In the second bin, 
$23.0<\Iacs<24.0$~mag, there is clear bimodality in \gIacs, with the red peak being
dominant. For $24.0<\Iacs<25.0$~mag, the blue peak becomes dominant, and this is true to an even
greater extent for the faintest magnitude bin of $25.0<\Iacs<26.0$~mag.

Figure~\ref{fig:hist2} shows the corresponding histograms of \IHacs\ color
using the same magnitude bins as in Figure~\ref{fig:hist1}. The samples are smaller
because of the smaller field of WFC3/IR, but again we find that the color
distribution appears broad, unimodal, and red for the brightest magnitude bin. Although
any bimodality is much less evident than in \gIacs, the \IHhst\ histogram for the
$23.0<\Iacs<24.0$~mag range is skewed towards the red, while the histograms for the
faintest two plotted magnitude ranges become progressively more skewed towards the
blue. This is qualitatively similar to what is observed for \gIacs, and it is consistent
with the striking ``mean tilt'' in the \IHhst\ versus \Iacs\ color-magnitude relation
(Figure~\ref{fig:cmdtilt}), for which the colors become bluer in \IHhst\ at fainter
magnitudes. 

\begin{figure*}
\begin{center}
\includegraphics[width=15.2cm]{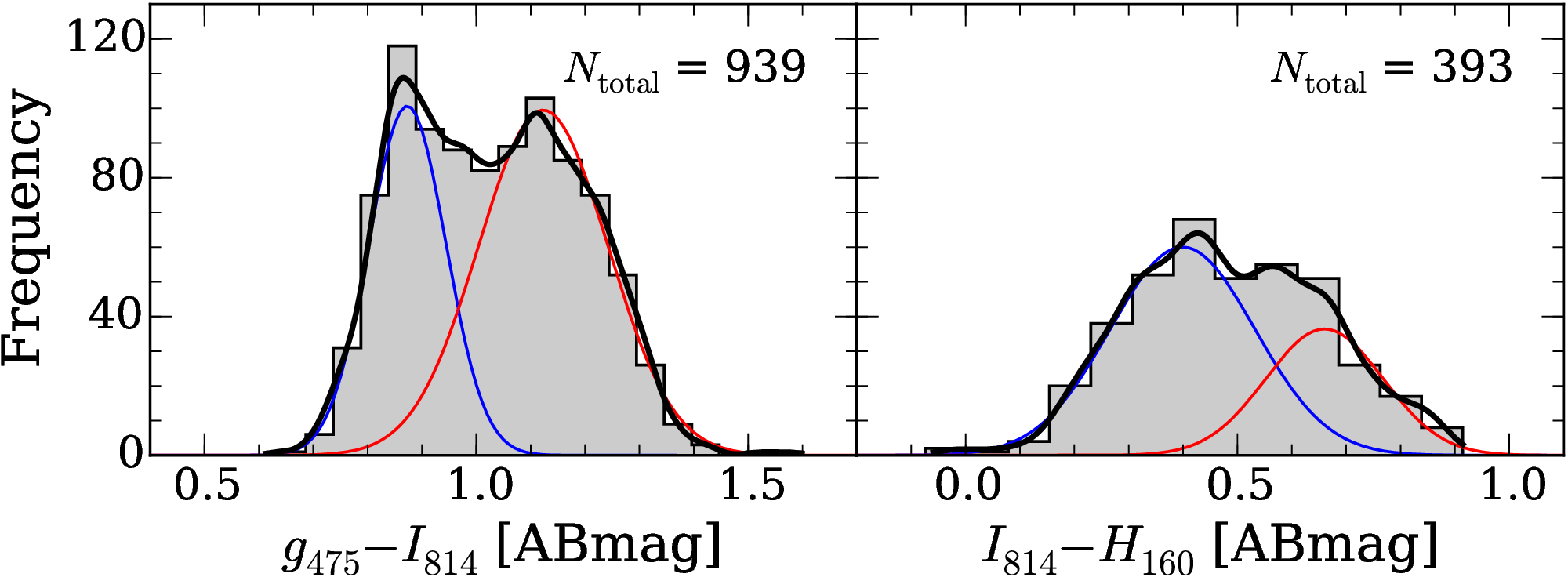}
\caption{
Histograms of \gIacs\ and \IHacs\ colors for GC candidates within the magnitude 
range of $23.0 < \Iacs < 25.0$~mag. The black thick solid curve is 
the nonparametric density estimate constructed with a Gaussian kernel.
We also plot the GMM double Gaussian model components for 
the heteroscedastic case with blue and red solid curves. 
The corresponding GMM analysis results are provided in Table~\ref{tab:GMM1}.
\label{fig:hist3}}
\end{center}
\end{figure*}

In order to quantify the visual impressions given by Figures~\ref{fig:hist1}
and~\ref{fig:hist2}, we ran the GMM code on the \gIacs\ and \IHhst\ color distributions
of the GC candidates in the various magnitude ranges shown in those figures.
Table~\ref{tab:GMM1} summarizes the results of these GMM analysis runs, as well as the
results for the broader magnitude range of $23.0<\Iacs<25.0$~mag, for which the \gIacs\ 
and \IHhst\ histograms are displayed in Figure~\ref{fig:hist3}. 
The bimodality in \gIacs\ is significant for all the magnitude ranges explored in
Table~\ref{tab:GMM1} except for the brightest; all the other bins have $p(\chi^2)<0.01$,
indicating less than 1\% probability of the color data being drawn from a single
Gaussian model, rather than the best-fit double Gaussian model with the tabulated means $\mu_1,
\mu_2$ and dispersions $\sigma_1, \sigma_2$ and with the fraction of objects in the
second (red) Gaussian given by $f_2$. The evidence for bimodality is stronger if the
tabulated $D$, the separation between the Gaussians in units of the quadrature sum of
their dispersions, is significantly $>2$, and if the kurtosis of the distribution 
$\kurt<0$ (see Muratov \& Gnedin 2010 and Blakeslee \etal\ 2012).
The optical bimodality is especially pronounced, and the double Gaussian model parameters best
constrained, within the $23.0<\Iacs<25.0$~mag range.

\begin{figure*}
\begin{center}
\includegraphics[width=15.2cm]{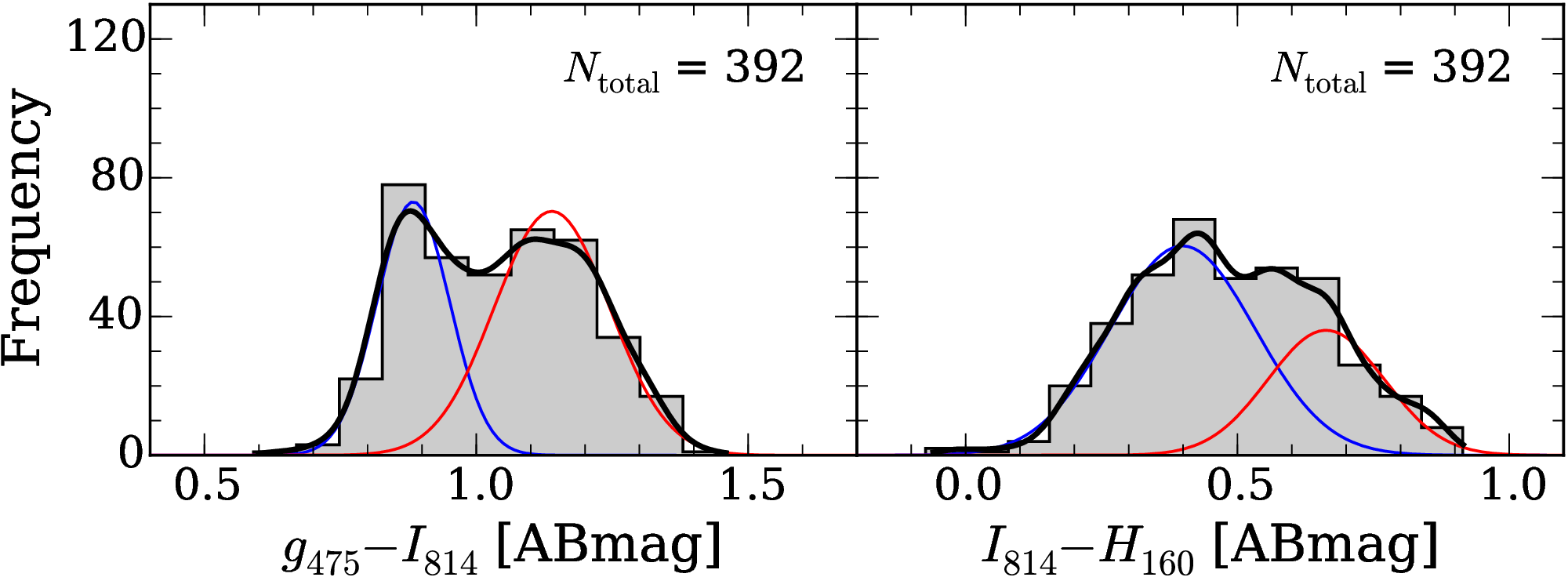}
\caption{
Same as Figure~\ref{fig:hist3} but for the cross-matched subsample of GC candidates with
both ACS/WFC F475W and WFC3/IR F160W data. 
The GMM analysis results are in Table~\ref{tab:GMM2}. 
\label{fig:hist4}}
\end{center}
\end{figure*}

For the \IHhst\ color index, the bimodality is only significant at the $>2\sigma$ level,
$p(\chi^2)<0.05$, for the $23.0<\Iacs<25.0$~mag range. Interestingly, however, for this
magnitude range, the GMM code gives $f_2 = 0.624\pm0.055$ for \gIacs, but $f_2 =
0.330\pm0.144$ for \IHhst. Thus, although the bimodality is significant in this magnitude
range for both \gIacs\ and \IHhst, the preferred ratios of red/blue GCs differ at the
$\sim2\sigma$ level. This is similar to the result for NGC\,1399, the cD galaxy in the
Fornax cluster, for which Blakeslee \etal\ (2012) found significantly different
bimodalities in \gIacs\ and \IHhst, resulting from the nonlinear relation between these
two color indices. However, it should be noted that although the magnitude range is the
same, the \gIacs\ sample has 939 objects while the \IHhst\ sample has only 393 objects
because the WFC3/IR field of view is smaller; 
it is not clear if the difference in color bimodalities is significant or not because the samples are different.

Figure~\ref{fig:hist4} shows the histograms for the cross-matched subsample of 392 GC
candidates in the $23.0<\Iacs<25.0$~mag range having both \gIacs\ and \IHhst\
colors. (There was one object in the sample of GC candidates with \IHhst\ colors that
was not included in the sample with \gIacs\ colors.) The optical \gIacs\ color is clearly
bimodal, while 
the separation remains less clear for \IHhst. Table~\ref{tab:GMM2} presents the GMM
analysis results for the \gIacs\ and \IHhst\ distributions of this homogeneous
cross-matched sample. We include both the homoscedastic (common dispersion,
$\sigma_1=\sigma_2$) and heteroscedastic ($\sigma_1\neq\sigma_2$) cases. 
For the heteroscedastic case, the preferred bimodal decompositions again differ
significantly, with $f_2 = 0.604\pm0.083$ and $f_2 = 0.324\pm0.128$ for \gIacs\ and
\IHhst, respectively. On the other hand, for the homoscedastic case, the GMM code finds
$f_2 = 0.493\pm0.031$ and $f_2 = 0.436\pm0.054$ for 
\gIacs\ and \IHhst, respectively. Thus, if the color dispersion for the blue and red GC
components are forced to be the same, then the bimodal decompositions for \gIacs\ and
\IHhst\ are consistent. However, the heteroscedastic GMM results imply that the 
dispersions differ significantly, at least for the purely optical \gIacs\ color,
with the blue peak being significantly narrower; the same result has been found for other
massive galaxies (e.g., Peng \etal\ 2006, 2009; Harris \etal\ 2016).

The heteroscedastic GMM results for \IHhst\ in Tables~\ref{tab:GMM1} and~\ref{tab:GMM2}
indicate that the color dispersion is slightly larger for the blue component than for the
red component, the opposite of what we find for \gIacs. Exploring this issue in more
detail, we found that the dispersion of the blue component in \IHhst, as well as the
blue:red ratio, was sensitive to the presence of a small number of GC candidates with the bluest
\IHhst\ colors. Table~\ref{tab:GMM3} reports the results for heteroscedastic GMM tests
when the two and four bluest GCs in \IHhst\ are removed from the sample. For instance,
when the blue limit is changed by $+0.15$~mag in \IHhst, reducing the sample size from
392 to 388, the blue component of the GMM decomposition becomes significantly narrower
and the preferred red fraction goes from $f_2=0.32\pm0.13$ to $f_2=0.54\pm0.13$, which is
consistent with the $f_2=0.61\pm0.08$ found for \gIacs. Thus, unlike the case for
NGC\,1399 (Blakeslee \etal\ 2012), the GMM decompositions of the matched sample are
consistent for the optical \gIacs\ and optical-IR \IHhst\ colors, after removing a few of
the bluest objects. However, we emphasize that 
the GMM decomposition is not very robust for \IHhst, mainly
because the separation $D$ of the blue and red components is not significantly greater than two,
and thus any bimodality is difficult to quantify.

\subsection{The Color-Color Relation}
\label{subsec:ccr}

We now explore the relation between the sets of color measurements presented in
the previous sections. 
As discussed by Blakeslee \etal\ (2012), optical and mixed optical-NIR color indices probe 
different spectral regions and therefore different properties of unresolved stellar systems. 
The \gIacs\ color is sensitive to the main sequence turnoff (which depends on the
turnoff mass, and thus on age), the
horizontal branch morphology (which behaves nonlinearly with metallicity and also
depends on age; Lee \etal\ 1994; Dotter \etal\ 2010), and the temperature of the red
giant branch. 
The \IHhst\ color is primarily sensitive to the temperature of the red giant branch, which
mainly depends on metallicity (e.g., Bergbusch \& VandenBerg 2001; Dotter \etal\ 2007).
Assuming similarly old ages for all the GCs, the form of the relation between different color
indices reveals whether the colors behave differently as a function of
metallicity, and thus can provide information on the color-metallicity relations.

\begin{figure}
\begin{center}
\includegraphics[width=1.00\linewidth, trim=0.cm 0 0 -0.3cm]{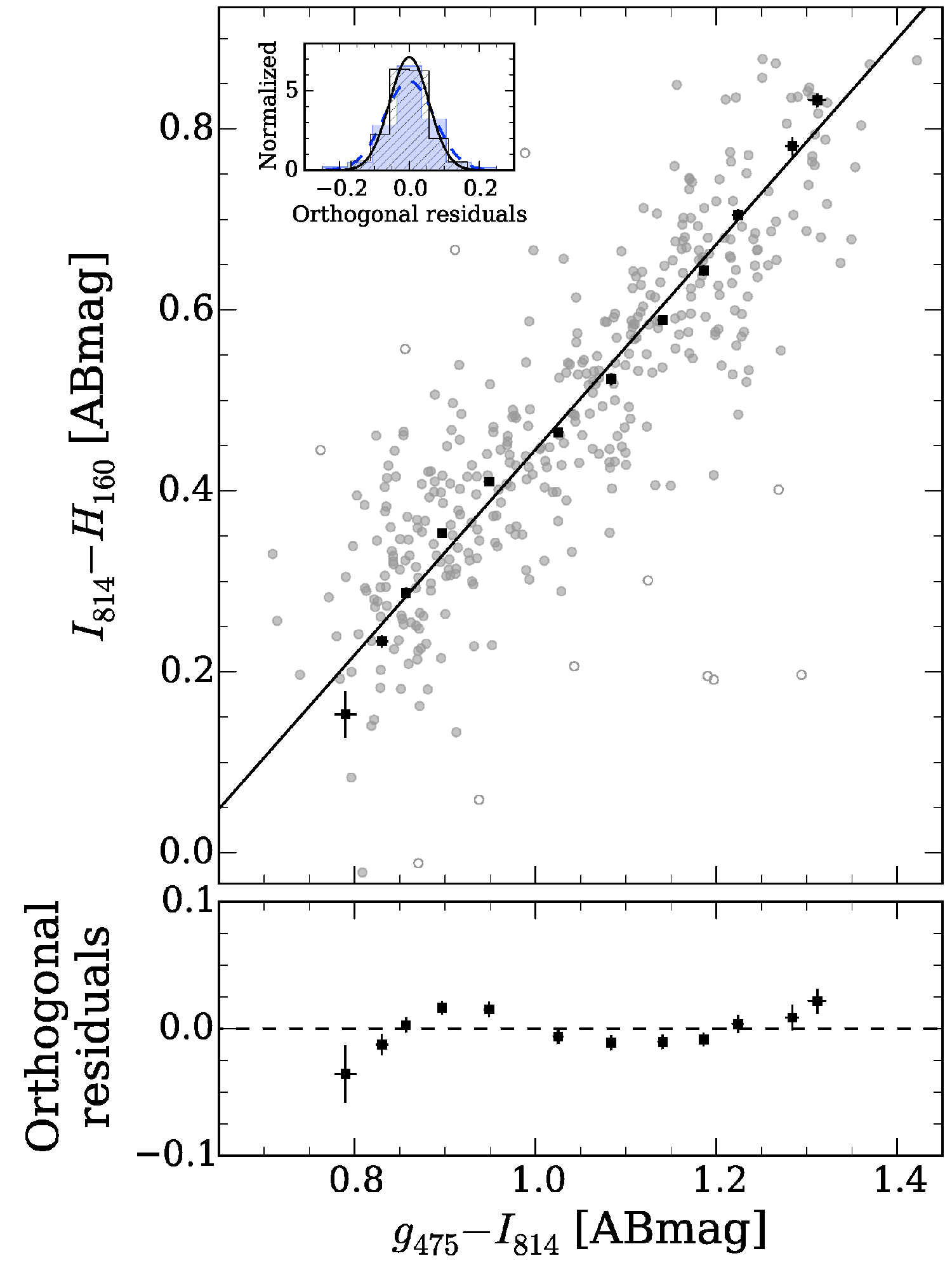}
\caption{
Optical-NIR \IHacs\ vs. purely optical \gIacs\ color.
The upper panel shows the color-color plane for individual GCs (solid and open 
gray points) and the modal (most probable) values within twelve bins
(dark points with error bars) along with the bisector line (solid black line).
Outlier rejection was done by an iterative $3\sigma$-clipping procedure based 
on minimizing the orthogonal distances of the data points from the linear 
bisector fit assuming normal distributions, as shown in the inset box 
(see Section~\ref{subsec:ccr} for details). The filled gray circles are 
the final data points after outlier rejection, while the open circles 
show rejected outliers. The blue dashed curve in the inset box is a normal distribution 
with $\sigma_{\rm initial}=0.071$ for the light-blue shaded histogram before clipping. 
The black solid curve in the inset box is a normal distribution with
$\sigma_{\rm final}=0.056$~mag for the black hatched histogram after clipping.
The bin width along the bisector line was chosen to be $3\sigma_{\rm final}$,
and the bin spacing is half of the bin width.  The lower panel shows the orthogonal
deviations (black squares with error bars) of the twelve binned data points with
respect to the best-fit line in the upper panel.
The plotted points indicate the modal values within each bin.
\label{fig:ccr1}}
\end{center}
\end{figure}

Figure~\ref{fig:ccr1} shows \IHhst\ as a function of \gIacs\ for
the matched sample of GC candidates in NGC\,4874. We plot only objects in the
$23.0<\Iacs<25.0$~mag range, where the color errors are small and the optical
bimodality is most pronounced. The figure shows the best-fit bisector line, i.e., the
linear relation that minimizes the orthogonal squared deviations, with 3-$\sigma$
clipping; the clipped points are plotted as open circles.
The inset box in Figure~\ref{fig:ccr1} shows that a normal distribution with dispersion 
$\sigma\approx0.06$~mag provides a good representation of the orthogonal color residuals
after clipping.
In order to study deviations from a linear color-color relation, we grouped the data into
twelve bins along the bisector line; the black squares in the upper panel of
Figure~\ref{fig:ccr1} show the modal locations within each bin, and the lower panel
shows the orthogonal deviations of these bins from the linear relation. Eight of the
twelve points deviate significantly from the linear relation, following an inflected,
or ``wavy,'' locus at least qualitatively similar to the results found in other studies of
the relations between optical and optical-NIR GC colors using high-quality photometric
data sets (e.g., Blakeslee \etal\ 2012; Chies-Santos \etal\ 2012; Cantiello \etal\ 2014). 

\begin{figure}
\begin{center}
\includegraphics[width=1.00\linewidth]{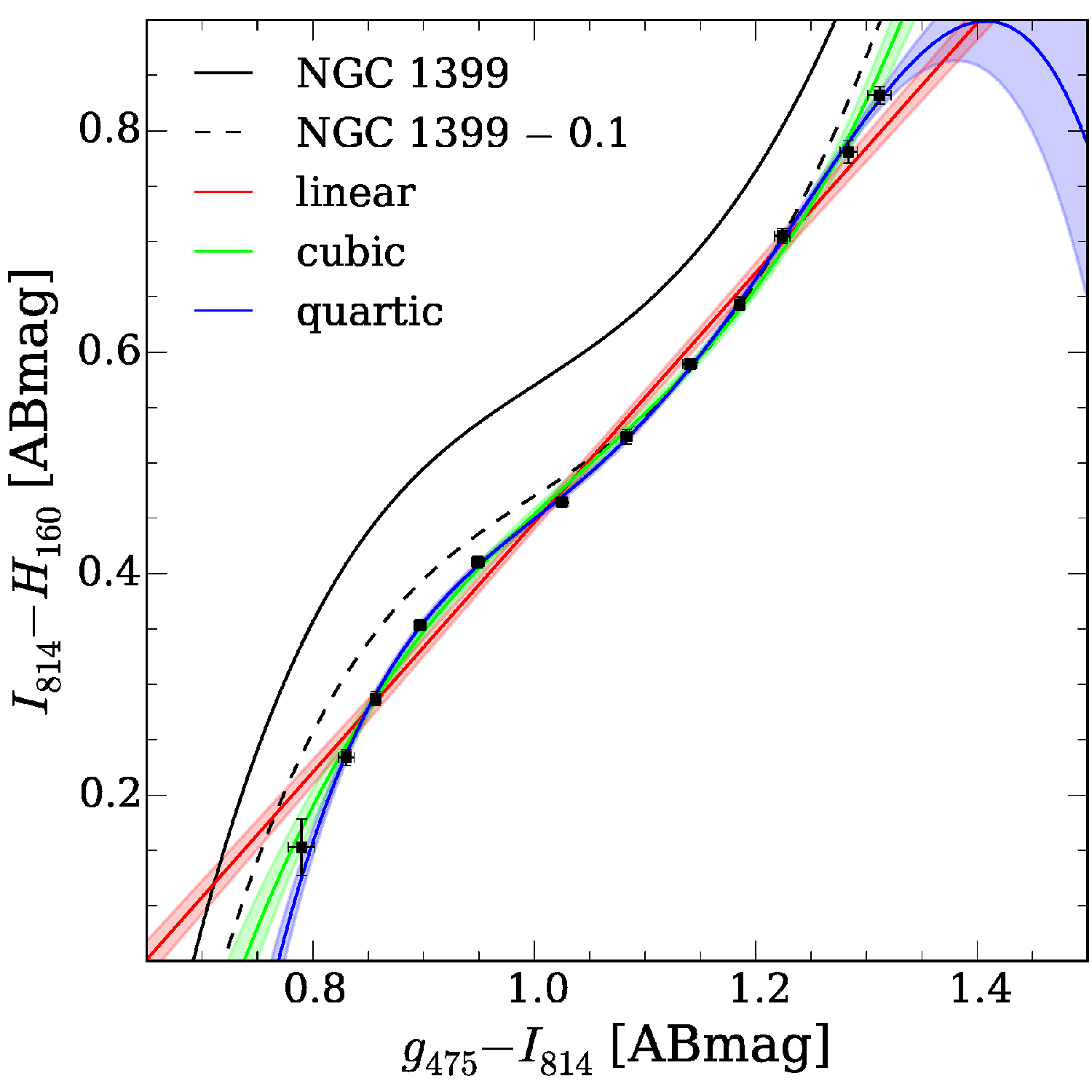}
\caption{
\IHacs\ vs. \gIacs\ color-color relation with weighted fits that minimize
the squared orthogonal distances of the binned data points (black squares,
same as plotted in the upper panel of Figure~\ref{fig:ccr1}). The coefficients of 
each fit are given in the text.  The 1-$\sigma$ ranges of the various fits are indicated
with shaded regions. The black solid curve shows  
the relation from Blakeslee \etal\ (2012) and the black dashed curve indicates 
the relation shifted by $-0.1$~mag in \IHacs\ (see~text).
\label{fig:ccr3}}
\end{center}
\end{figure}

The binned modal values of the relation between \IHhst\ and \gIacs\ are again shown in
Figure~\ref{fig:ccr3}, along with several different polynomial fits. Similar to our previous
work (Blakeslee \etal\ 2012), the fits are robust orthogonal regressions, weighted by
the uncertainties on the individual binned values; we also show the 1-$\sigma$
uncertainty regions around the fits. The equations for the plotted linear, cubic, and
quartic fits are, respectively:
\begin{eqnarray}
\IHacs &=& (-0.68\pm0.04) + (1.13\pm0.04)x \,, \label{eq:gIvsIHlinear}\\
\IHacs &=& (-7.28\pm1.53) + (20.60\pm4.45)x \nonumber\\
  &+& (-18.91\pm4.28)x^{2} + (6.05\pm1.36)x^{3}\,,\\
\IHacs &=& (-37.11\pm6.17) + (136.11\pm23.59)x \nonumber\\
  &+& (-185.27\pm33.60)x^{2} + (111.68\pm21.12)x^{3} \nonumber\\
  &+& (-24.95\pm4.94)x^{4}\,, 
\end{eqnarray}
where $x\equiv\gIacs$.
Both the cubic and quartic polynomials provide statistically acceptable (within
${\sim\,}1\sigma$) descriptions of the data over the applicable domain 
$0.8\lta\gIacs\lta1.3$, while the linear fit is rejected with more than
99.9\% probability. Thus, the relation is nonlinear to a high degree of significance.

The quartic fit derived for globular clusters in the Fornax cD galaxy NGC\,1399 
(Blakeslee \etal\ 2012) is also plotted in Figure~\ref{fig:ccr3}. 
Unlike in the present analysis, the 3-pixel aperture magnitudes from that study were not
aperture corrected. This is a small effect for \gIacs\ because both \gacs\ and
\Iacs\ are measured on ACS data with the same pixel scale and similar PSFs; thus, the
differential aperture correction between the two bands is small. However, it is a much
larger effect for \IHhst\ because the stacked WFC3/IR images have twice the pixel scale
of the stacked ACS images, and GCs are significantly resolved at the 20~Mpc distance of
the Fornax cluster. Assuming King model profiles with the range of half-light radii for
GCs in the ACS Fornax Cluster Survey (Masters \etal\ 2010), we find that the correction in
\IHhst\ would be in the range of $\sim\,$0.05 to $\sim\,$0.1~mag. Figure~\ref{fig:ccr3}
shows that shifting the uncorrected 3-pix aperture color relation for NGC\,1399 by
0.1~mag provides an approximate (though not statistically acceptable) match to the
NGC\,4874 relation. This remaining disagreement may result from still larger
differential aperture effects at the blue end, where GCs tend to have larger sizes 
(Jord\'{a}n \etal\ 2005; Masters \etal\ 2010), and/or intrinsic differences in the
color-color relations and the underlying color-metallicity relations. Usher \etal\ (2015) 
found that there are significant differences in the $g{-}i$ color-metallicity
relations for different galaxies. We plan to address this issue fully in a future paper
presenting the \hst\ optical-IR colors of GCs in a larger sample of Virgo and Fornax
cluster galaxies, including detailed modeling of the differential aperture effects at
these more nearby distances. For now, we conclude that the relation between \gIacs\ and
\IHhst\ for GCs in NGC\,4874 appears to have less extreme curvature than our previously
published relation for NGC\,1399, but the deviation from a purely linear relation
remains highly significant.

\subsection{Radial Distributions}
\label{subsec:radialdist}

The spatial distributions of GCs around galaxies and within galaxy clusters can provide
information on the buildup of galaxy halos and cluster dynamical histories (e.g., 
Moore \etal\ 2006; Mackey \etal\ 2010; Lee \etal\ 2010; Keller \etal\ 2012).
Evidence for sizable populations of IGCs, objects bound to the
overall cluster potential rather than any individual galaxy,
has been found in several massive galaxy clusters, including Virgo (Lee \etal\ 2010; Durrell \etal\ 2014),
Coma (Peng \etal\ 2011), Abell 1185 (West \etal\ 2011), and Abell 1689
(Alamo-Mart\'{\i}nez \etal\ 2013).
Numerical studies (Bekki \& Yahagi 2006; Smith \etal\ 2013, 2015;
Mistani et al.\ 2016) find that dwarf
galaxies can lose substantial fractions of their GC systems to the larger cluster
environment, but they come to varied conclusions regarding whether the bulk of the
IGC population results from stripping of dwarfs or the outskirts of
more massive galaxies.
Peng \etal\ (2011) found an extensive population of IGCs in the center of the Coma cluster.
Because these IGCs showed a significant tail of red GCs comprising roughly 20\% of the
population, the authors concluded that a sizable fraction of the IGCs originated in
massive galaxies, rather than from disrupted dwarfs.

\begin{figure}
\begin{center}
\includegraphics[width=1.0\linewidth]{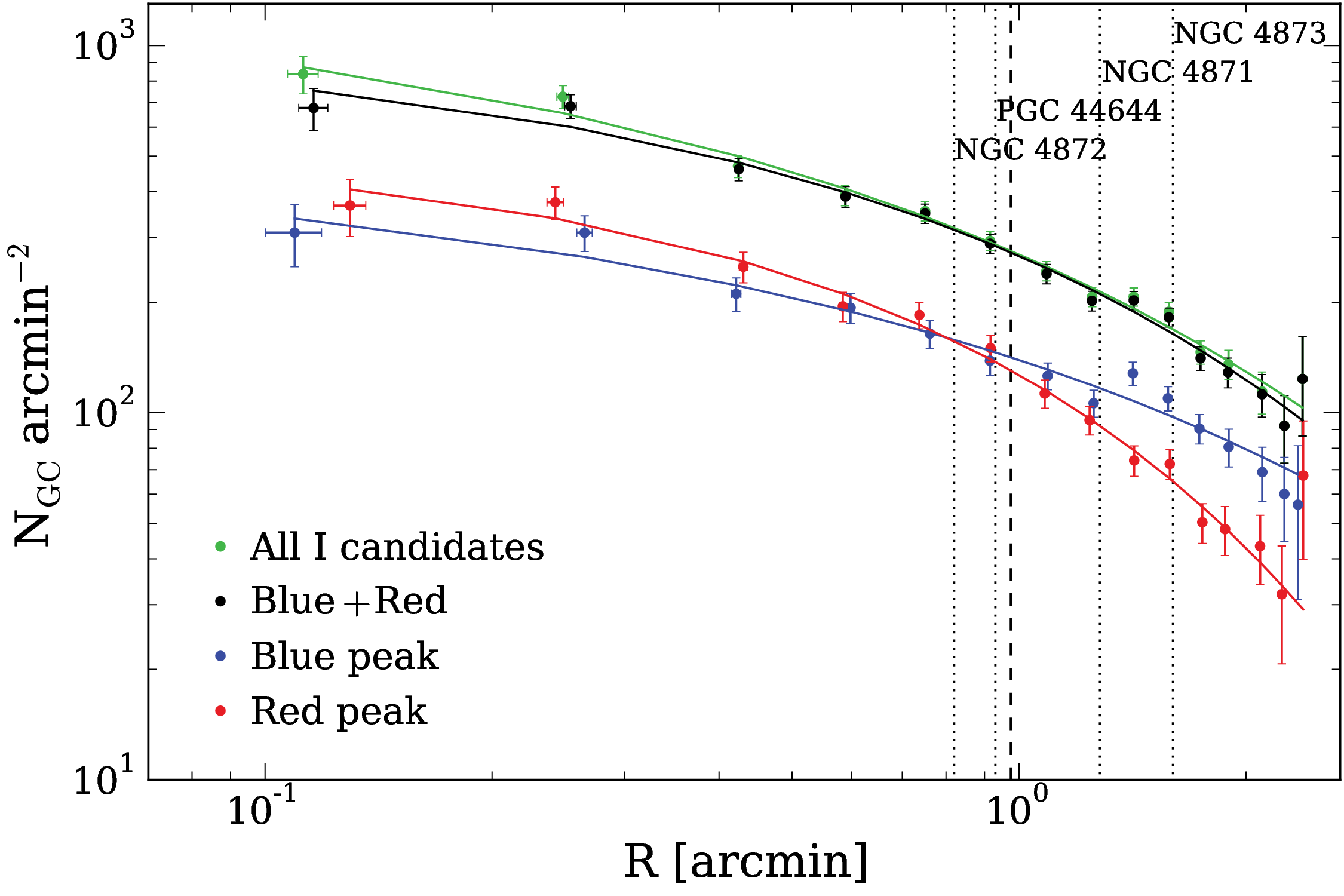}
\caption{
Number densities of GC candidates 
within the magnitude range $23.0 < \Iacs < 26.0$~mag as a function of 
galactocentric radius $R$ from NGC\,4874. 
The color-coded points represent number density estimates 
within fixed $10\arcsec$ annuli for 
different GC subsamples as described in the legend.
The values are plotted at the locations of the 
median galactocentric radius of the GCs within each bin.
The horizontal error bars represent the standard deviations of the
radial positions within each annulus.  
The vertical error bars indicate the Poisson errors on the
counts, i.e., $\sqrt{N}/$(effective area), within each annulus.
The best-fit \sersic\ profiles for each subsample are displayed as 
color-coded curves. The labeled vertical dotted lines indicate the distance 
from NGC\,4874 to the four nearest surrounding galaxies. The half size of the 
apparent isophotal diameter $\rm D_{25}$ to $\mu_{B} = 25.0$~mag/arcsec$^2$ 
(de Vaucouleurs \etal\ 1991) of NGC\,4874 is shown as a vertical dashed line.
\label{fig:blueredrad}}
\end{center}
\end{figure}

In order to quantify the spatial distribution of the GCs, and differences between the
red and blue subcomponents, we analyzed the projected surface number density profiles of
the ACS GC candidates in the $23.0<\Iacs<26.0$~mag range 
as a function of galactocentric radius $R$.
(Note that we have not integrated the observed counts over an assumed
GCLF, in contrast to Peng et al. 2011.)
From the center of
NGC\,4874 to $R=150\arcsec$, the GCs were binned within fixed radial annuli
of $10\arcsec$ width.
The number of GCs within each annulus was normalized by the effective area
of the annulus to get the number densities, and these are plotted against $R$ in
Figure~\ref{fig:blueredrad}, along with their Poisson-based uncertainties.
We fitted the number densities with the commonly
used \sersic\ (\sersic\ 1963) profile: 
\begin{equation}
N_{\rm GC}(R) \,=\, N_{\rm e} \, \exp\left\lbrace -b_{n}\left[ \left(\frac{R}{R_{\rm 
e}}\right)^{1/n} - \,1\right] \right\rbrace\,, 
\end{equation}
where $N_{\rm e}$ is the projected number density at the effective radius $R_{\rm e}$, 
$n$ is the \sersic\ index, and the constant $b_{n} = 1.9992n - 0.3271$ 
(Graham \& Driver 2005). We have not fitted a background level
because the radial coverage of our data is not wide enough to estimate it. 
Based on Peng \etal\ (2011), the expected background of point-like sources
at \hst/ACS resolution over this magnitude range is more than an order of magnitude
below the number densities in our outermost bins (and more than two orders of magnitude
below the innermost bins), even when the GCs are split into blue and red groups.

The \sersic\ fits to the full radial ranges are shown in Figure~\ref{fig:blueredrad}; the
reduced $\chi^2$ values for these fits are typically $\sim0.9$, indicating that the fits
provide reasonable descriptions of the data over these radial ranges.
For the full color-selected sample of GCs with $0.5<\gIacs<1.6$
(plotted as black points in Figure~\ref{fig:blueredrad}), we derive a \sersic\ index
$n_{\rm B+R} = 1.5\pm0.3$ with an effective radius of 
$R_{\rm e,B+R} = 4\farcm2\pm1\farcm5$, corresponding to $122\pm44$~kpc.
Peng \etal\ (2011), using the shallower ACSCCS data, but covering a larger area of the
Coma cluster, found 
$n= 1.3\pm0.1$, $\Ref = 2\farcm2\pm0\farcm1$, corresponding to $62\pm2$~kpc.
Our value of $n$ agrees closely with this ACSCCS value, while our \Ref\ is larger by a
factor of $2.0\pm0.7$, or a $1.4\sigma$ discrepancy.
Because our \gacs\ imaging is significantly less deep than \Iacs, and could
potentially affect the completeness of innermost bins, we
also fitted the number densities for all the \Iacs-selected GC candidates over the same 
$23.0<\Iacs<26.0$~mag range, but without matching to the \gacs\ detections. The
resulting densities are represented by green points in Figure~\ref{fig:blueredrad}; as
expected, they only differ at the 1-$\sigma$ level for the innermost point.
Our \sersic\ fit to this sample of ``all'' \Iacs-selected GCs gives
$n_{\rm all} = 2.0\pm0.4$, $R_{\rm e,all} = 3\farcm0\pm0\farcm7$, or $85\pm21$~kpc.
For this case, \Ref\ agrees to better than 1.1$\sigma$, while $n$
differs by 1.7$\sigma$. Given the differences in depth and area for these fits, the
level of agreement with the ACSCCS study is reasonable.

Peng \etal\ (2011) chose not to fit the blue and red GC components individually;
this was in 
part because the separation between the two color components varied with position
over the large area that they studied.
For instance, they found that the blue peak of the GC population within 
$R<50$~kpc of NGC\,4874 occurred between the locations of the blue and red peaks in the
GC color distribution at larger radius.  Since our deeper imaging data are limited 
to this one central pointing, we here examine the radial distributions of the blue and
red GCs separately, using the color at the local minimum (approximately $\gIacs=1.0$) of the
nonparametric kernel density estimate shown in Figure~\ref{fig:hist3}
to divide the GCs into ``blue'' and ``red'' subpopulations.
Fitting each of these color components with \sersic\ profiles, we find 
$n_{\rm B} = 1.9\pm0.7$, 
$R_{\rm e,B}=7\farcm0\pm6\farcm3$, 
$n_{\rm R} = 1.2\pm0.2$, 
$R_{\rm e,R}=1\farcm6\pm0\farcm2$
for the blue and red GCs, respectively.

The large uncertainty for the blue peak subpopulation is mainly due to the fact that the
effective radius is apparently much larger than our field of view. The large \Ref\ for
the blue GC distribution is likely related to the very extended, mainly blue 
IGC population in Coma (Peng \etal\ 2011). However, it is also related to the
mainly blue populations of GCs around the lower luminosity, but
still bright, elliptical galaxies in this field. Figure~\ref{fig:blueredrad} indicates
the radial locations of several neighboring galaxies. The density of blue GCs appears to
jump upward near the radii where NGC\,4871 and NGC\,4873 are located.
While these galaxies are bright enough to harbor some red GCs, the mean
color of these GCs
would be significantly bluer than those of NGC\,4874
(e.g., Peng et al.\ 2006), meaning that some of the smaller neighbors'
red GCs would fall within the 
range of the blue GCs for NGC\,4874. Moreover, the reddest GCs in these
neighboring galaxies would be restricted to small
galactocentric radii, where the completeness of our \gacs\
data suffers. The GCs at larger radii within these galaxies are overwhelmingly blue.

Because there is an inherent covariance between $n$ and \Ref\ for \sersic\ model fits
when the measurements do not extend clearly beyond \Ref, we have refitted the various
GC samples with $n$ fixed at~2.0. To avoid concern over possible incompleteness near the
bright galaxy center, we also omit the central radial bin for these fits.
With these constraints, we then find:
$R_{\rm e,all} = 4\farcm2\pm0\farcm3$, 
$R_{\rm e,B+R} = 4\farcm2\pm0\farcm3$, 
$R_{\rm e,B} = 7\farcm2\pm1\farcm2$, and
$R_{\rm e,R} = 2\farcm6\pm0\farcm2$. 
Thus, when $n$ is fixed and the central bin is omitted, matching with the
\gacs\ detections and limiting the color range does not change the resulting profile.
However, we continue to find that the effective radius of the radial distribution of the
blue GCs is significantly larger than that of the red GCs. Again, this is in part due to
the contribution of blue GCs from NGC\,4871, NGC\,4873, and other galaxies.
It is also consistent with the radially declining fraction of red GCs found 
by Peng \etal\ (2011) over a larger area of the Coma core, and many other studies 
that find the red GCs are more concentrated in giant ellipticals and within galaxy
clusters (e.g., Faifer \etal\ 2011; Durrell \etal\ 2014).

\subsection{Two-Dimensional Spatial Distribution}
\label{subsec:spatialdist}

\begin{figure}
\begin{center}
\includegraphics[width=1.0\linewidth, trim=0 0 0 0.4cm]{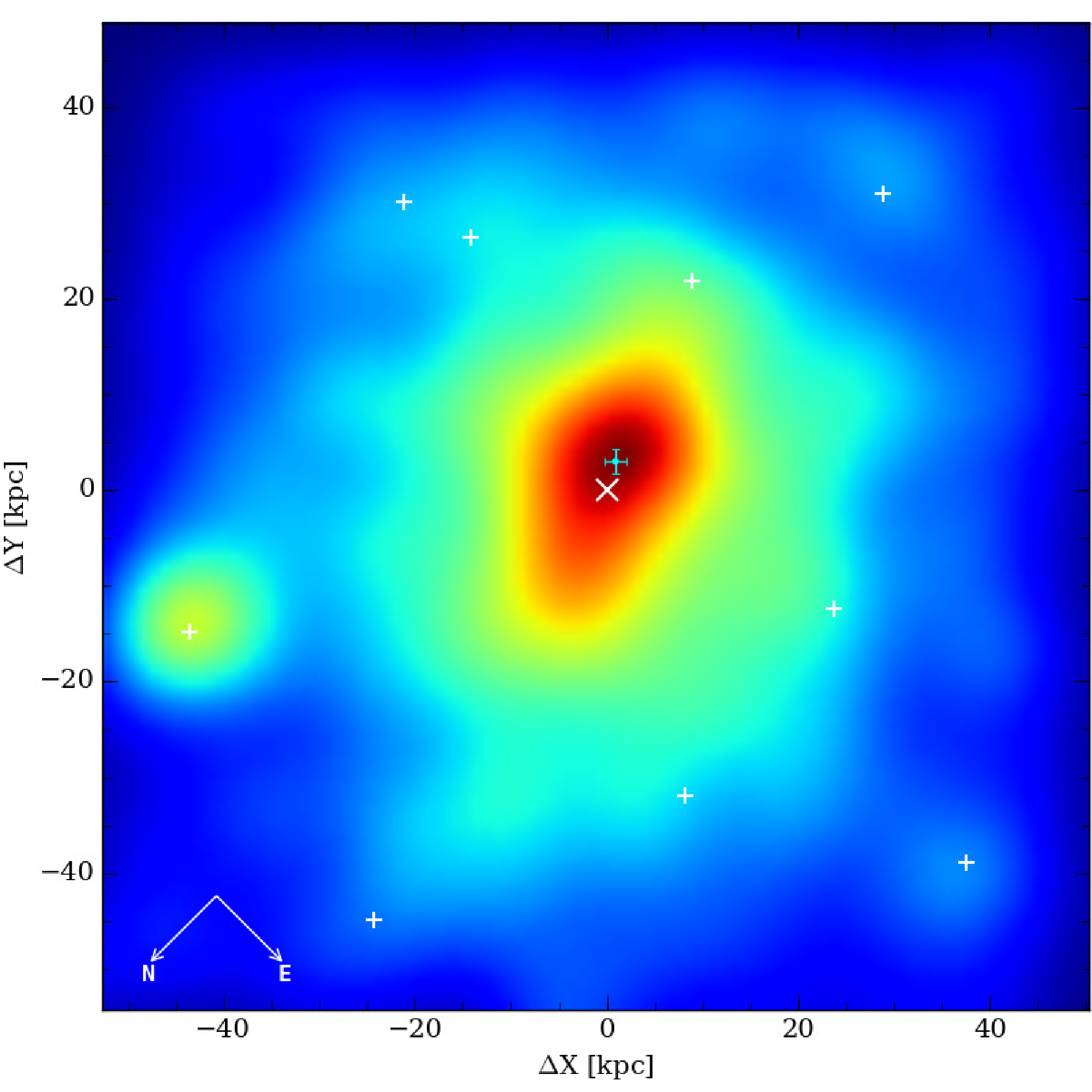}
\caption{
Example smoothed two-dimensional surface density map of the spatial distribution of GCs 
from the $I$-band photometry in the range $23.0 < \Iacs < 26.0$~mag (color 
indicates surface number density of GCs). 
The grid size of the two-dimensional histogram is 50 pix with a Gaussian 
smoothing kernel of $\sigma=4$ grid spacings for this example map; many other grid sizes
and smoothings were explored.
The white $\times$ symbol marks the center position of NGC\,4874 from 
IRAF {\tt ELLIPSE} fitting and the cyan point with error bars marks the average center of the 
GC population.
The small white $+$ symbols mark the locations of nine surrounding galaxies. 
\label{fig:spcolormap}}
\end{center}
\end{figure}

In order to investigate possible spatial differences between the distributions of the
stellar light and GCs in NGC\,4874, we constructed two-dimensional smoothed 
spatial number density maps of the GCs. 
For this purpose, we used GC candidates selected only from the \Iacs\ photometry, i.e.,
the ``All I candidates'' sample in Figure~\ref{fig:blueredrad}, 
since completeness may become a problem near the center of the galaxy for the ACS/F475W image. 
However, we have also repeated the full two-dimensional analysis using the matched F814W/F475W sample,
and the results do not change in any significant way.
To characterize the two-dimensional GC distribution, 
we divided the ACS field into two-dimensional grids with various grid sizes:
40, 50, 60, 70, 80, 90, and 100~pix on a side (recall the scale is 
0\farcs05~pix$^{-1}$ for our ACS imaging) 
and calculated the number density of GCs within each grid cell.
The resulting bi-dimensional histograms were then smoothed with Gaussian kernels 
with varying standard deviations of $\sigma = 2$, 3, and~4, in units of the grid spacing.
Surprisingly, the peaks of these smoothed two-dimensional GC density distributions
generally do not encompass the luminosity center of NGC\,4874, i.e., the GCs 
in the inner region of this field have an off-centered spatial distribution with respect
to NGC\,4874. An example smoothed GC surface density map (50~pix grid size with
$\sigma=4$ grid smoothing)
is shown in Figure~\ref{fig:spcolormap} with the locations of the ten brightest galaxies
marked. As evident in the figure, the peak of the GC density distribution is displaced
towards the south/southwest with respect to the center of NGC\,4874.

\begin{figure}
\begin{center}
\includegraphics[width=1.0\linewidth]{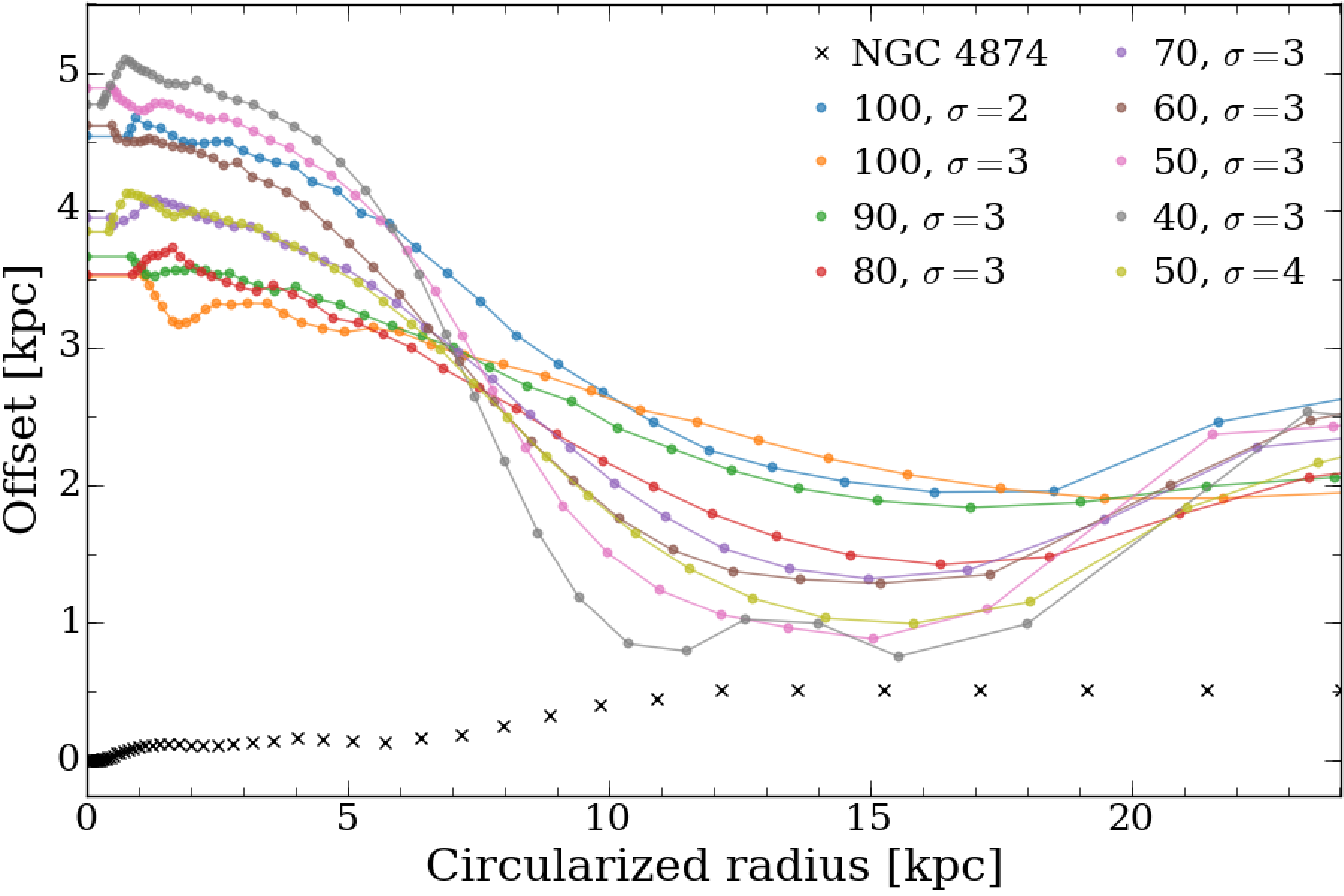}
\caption{
Offset distance between the peak of NGC\,4874's starlight and the centers of elliptical
isophotes of various circularized radii $r_{\rm cir}$.
The elliptical isophotes (or isodensity contours) are
fitted to multiple different 2-D representations 
(round points of various colors indicating the different 2-D 
binnings and smoothing scales)
of the GC number density distribution, as well as to
the galaxy light itself (black crosses). 
For the GC density distribution, the grid size (in pixels) of the
spatial binning and the Gaussian smoothing $\sigma$ (in units of grid spacing) are
indicated in the legend. The circularized radius is defined as 
$r_{\rm cir} = a\sqrt{1{-}\epsilon}$, where $a$ is the semi-major axis of each
elliptical isophote 
(or isodensity contour) and $\epsilon$ is its ellipticity.
Regardless of the factor-of-two range in grid size and smoothing scale, we
find that the central peak of the elliptical model of the GC density distribution is
offset by 3--5~kpc (6\arcsec\ to 10\arcsec) with respect to the central peak of
NGC\,4874 itself. 
At large radii, beyond $\sim10$~kpc, the center of the elliptical GC isodensity
contours approach to within about 1.5~kpc of the luminosity center of NGC\,4874.
\label{fig:offset}}
\end{center}
\end{figure}

To quantify the centroid of the GC distribution, we fitted elliptical isophotes
(representing GC number isodensity contours)
to the smoothed density maps using the IRAF {\tt ELLIPSE} task. The distance from 
the luminosity center of NGC\,4874 to the center of each ellipse is plotted in 
Figure~\ref{fig:offset} as a function of the circularized radius of each isophote
$r_{\rm cir} = a\sqrt{1-\epsilon}$, where $a$ is the semi-major axis and $\epsilon$ is 
the ellipticity of each ellipse. We estimated the statistical significance of 
the centroid offsets by bootstrap resampling of the GC spatial density distribution 
10,000 times before applying the two-dimensional smoothing. Figure~\ref{fig:offset} shows that,
regardless of the particular smoothing, the centroid of the GC density distribution is
displaced from the galaxy center by $4\pm1$~kpc (about~8\arcsec) towards 
the south/southwest from the center of stellar light distribution.
However, on larger scales, $r_{\rm cir}\gta10$~kpc, 
the centers of the GC isodensity contours approach within $\sim1$~kpc of the center of
the galaxy isophotes.
We note that Kim \etal\ (2013) also reported an offset (of $\sim3$~kpc) for the center
of the GC system around NGC 1399, the cD galaxy in the Fornax cluster.

Most likely this displacement in the centroid of the GC system is related to dynamical
interactions within this very rich environment. We note that NGC\,4889, the
brightest galaxy in the Coma cluster, is located approximately 200~kpc to the east, 
and thus does \textit{not} appear to be associated 
with the observed small offset of NGC\,4874's GC system. However, the offset
\textit{does} align closely with the direction towards NGC\,4872,
an S0 galaxy with a prominent bar.  At a separation
of only 0\farcm82, or 24~kpc, NGC\,4872 is the closest of the bright neighboring galaxies,
and its velocity (from NED) differs by only $17\pm4$~\kms\ from NGC\,4874.
Despite its luminosity, NGC\,4872 does not have an obvious GC system of its own
(unlike NGC\,4871 and NGC\,4873).
It would be interesting to explore through dynamical modeling if the observed offset of the
NGC\,4874 GC distribution could be related to dynamical interaction with NGC\,4872.

\begin{figure*}
\begin{center}
\includegraphics[width=0.9\linewidth]{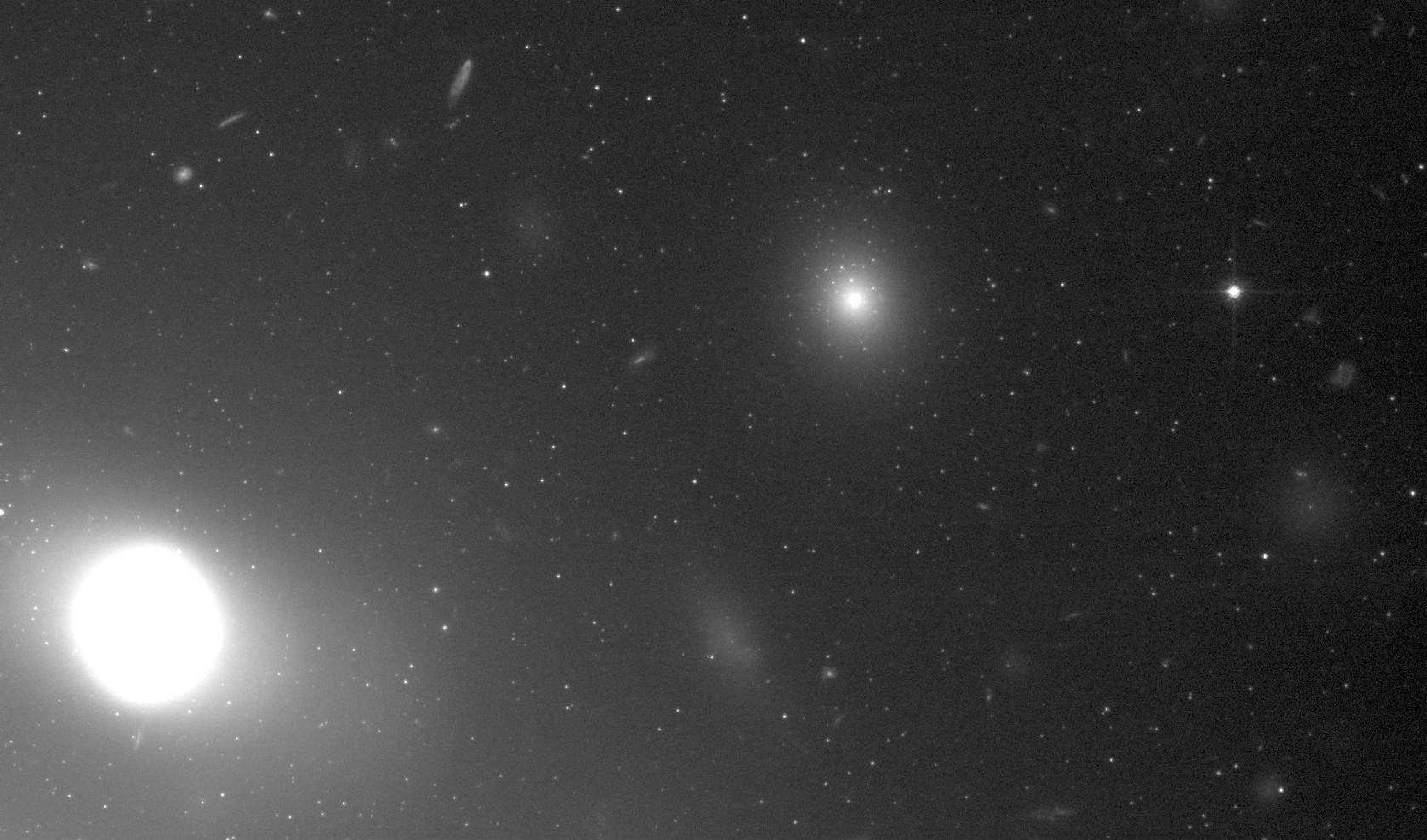}
\caption{
$84\arcsec{\,\times\,}50\arcsec$ ($\sim40$~kpc${\,\times\,}24$~kpc) section near the 
top~right corner of our deep F814W image (shown in full in Figure~\ref{fig:imgf814w}\,(a)).
The brightest galaxy here, near the lower left corner in this subimage, is the SB0
galaxy NGC\,4872 (labeled in Figure~\ref{fig:imgf814w}\,(a)).
The dwarf elliptical galaxy just to the upper right of center in this subimage is
SDSS\,J125935.18$+$275605.0.
It contains a sizable population of GCs with an asymmetric spatial distribution. 
Interestingly, this dE galaxy has a very high relative velocity of $-2660\pm160$~\kms\ with
respect to the cluster mean; this is about 2.5~times the cluster velocity dispersion. 
Several extremely diffuse galaxies, like those found in other recent
studies, are also evident in this field.
\label{fig:prettypic}} 
\end{center}
\end{figure*}

\subsection{A Dwarf Elliptical with an Asymmetrical GC System}

In the course of our analysis of the galaxy light distributions, we noticed one
particular dwarf elliptical (dE) galaxy 1\farcm47 from NGC\,4874 that seemed
relatively rich in GCs, but the GC distribution appeared strikingly asymmetrical. 
Searching its coordinates in NED, we found that 
the galaxy was catalogued in the Sloan Digital Sky Survey (SDSS) and
is designated SDSS\,J125935.18$+$275605.0; for convenience, we refer
to it hereafter as SDSS\,J125935. This is the faintest of the ten galaxies in our ACS images 
for which we performed isophotal modeling, and it lies near the top right corner of our ACS
field (outside our WFC3/IR imaging area); see the galaxy model panel in Figure~\ref{fig:imgf814w}.
Figure~\ref{fig:prettypic} shows an $84\arcsec\times50\arcsec$ cutout of the region around
this galaxy in our \Iacs\ image; the much brighter galaxy at lower left in this figure is
NGC\,4872, discussed above. Several fainter, more diffuse, objects in this field appear
similar to the extremely diffuse galaxies first systematically catalogued in the Coma
cluster by van Dokkum \etal\ (2015), and shown from deep Subaru imaging to be
ubiquitous throughout the Coma cluster (Koda \etal\ 2015). The image also shows that
the density of point sources around the dE SDSS\,J125935 is not symmetric
about the galaxy's center.

\begin{figure*}
\begin{center}
\includegraphics[width=0.85\linewidth]{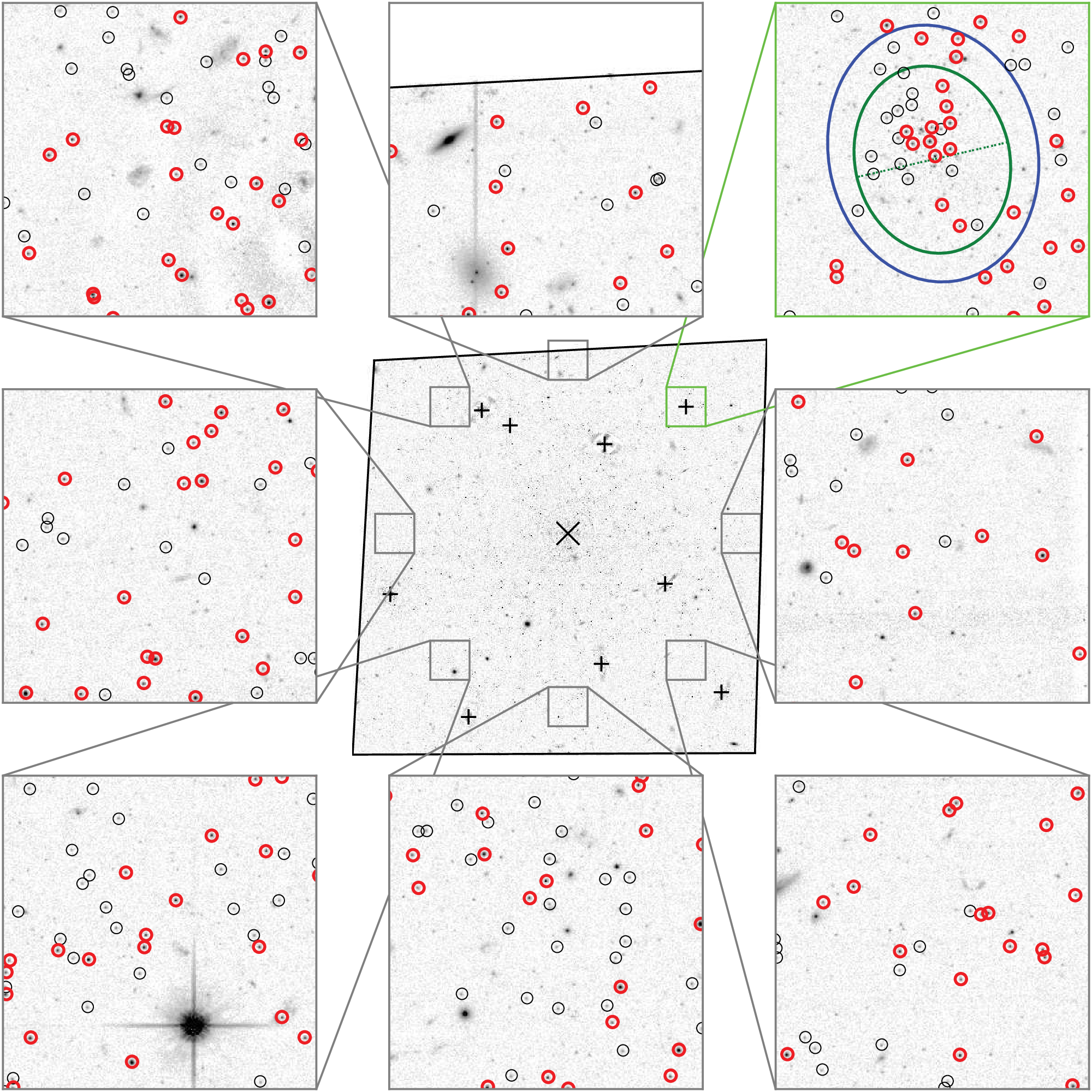}
\caption{
Spatial distributions of GCs around the dE galaxy SDSS\,J125935.18$+$275605.0
and in seven control fields at the same projected distance from the cD galaxy NGC\,4874.
Each of the small fields is $20\arcsec\times20\arcsec$ in size, except the top center
field, which is $20\arcsec\times15\arcsec$ because of its proximity to the image edge.
The dE has been subtracted via isophotal modeling; the larger (blue) ellipse
marks the outermost isophote for which we
were able to constrain the galaxy's ellipticity and position angle in our F814W imaging;
it has semi-major and semi-minor axes of 4.0~kpc and 3.2~kpc, respectively.
For comparison, the green ellipse indicates the $r=25$ mag~arcsec$^{-2}$ isophote catalogued
in the SDSS; the dashed line marks the minor axis of this ellipse.
Objects within red circles are GC candidates with $\Iacs<26$~mag, while those within black
circles have $26.0<\Iacs<26.9$~mag (see text). 
The GCs associated with SDSS\,J125935.18$+$275605.0 tend to fall towards one side
of the galaxy, in the direction opposite of NGC\,4874; 
the outer isophote also appears to be offset in this direction.
Given this dE galaxy's very high velocity relative to Coma, its proximity to the cluster
dynamical center, and its asymmetrically distributed GC system,
it may be passing through the core of
the cluster on its first infall and being stripped of much of its GC system.
\label{fig:dwarfgal}} 
\end{center}
\end{figure*}

Figure~\ref{fig:dwarfgal} further illustrates the spatial asymmetry of the GC distribution 
in a $20\arcsec\times20\arcsec$ box around this dE by comparing it to seven ``control'' 
fields of the same size and at the same radius with respect to the cD galaxy NGC\,4874. 
In the figure, GCs in the matched F475W+F814W sample with $\Iacs<26.0$~mag
are shown with red circles, while those with  $26<\Iacs<26.9$~mag (i.e., down to the
expected GCLF turnover) are shown with smaller black circles. Although we have found
that the completeness of the F475W detections is lower near the center of NGC\,4874
and the other bright ellipticals, the surface brightness of SDSS\,J125935 is low enough
that completeness is not a serious issue to this magnitude.
The dE itself has been subtracted using our isophotal model. The green ellipse
shows the $r=25$ mag~arcsec$^{-2}$ isophote from the SDSS, as
reported by NED, and the dashed line marks the minor axis of this ellipse.
The blue ellipse indicates the outermost isophote for which we
were able to constrain the galaxy's ellipticity and position angle from our deep F814W
image; it has a semi-major axis of 8\farcs25 and an ellipticity of 0.192. At the distance
of Coma, this translates to semi-major and semi-minor axes of 4.0 and 3.2~kpc, respectively.

For the GCs with $\Iacs<26.0$~mag, 9 of the 11 inside the green ellipse lie 
to one side of the minor axis; the probability of this occurring by chance is 6.5\%,
based on Monte Carlo tests.
However, the asymmetry is not limited to the brighter GCs; for those
with $\Iacs<26.9$~mag (red plus black circles), 17 of 23 lie to one side,
which has a random probability of 3.5\%.  Considering the larger blue
ellipse, 12 of 15 GCs with $\Iacs<26$~mag, and 23 of 31 GCs with $\Iacs<26.9$~mag,
lie to one side of the minor axis; these have random probabilities of 3.5\% and 1.1\%,
respectively. Thus, the asymmetry is significant with $\sim99$\% confidence.
It is evident from Figure~\ref{fig:dwarfgal} that the outermost galaxy isophote is also
offset slightly (centroid shift of 0\farcs4) in the same direction as the GCs.

We can estimate the size of the GC population in SDSS\,J125935 from the $31\pm5.6$ GC
candidates (the error is based on Poisson statistics) with $\Iacs<26.9$~mag within the blue
ellipse in Figure~\ref{fig:dwarfgal}.  This ellipse has an area of 172.9~arcsec$^2$, and
based on the density of GC candidates in the control fields, we would expect $14\pm4$ 
contaminants (mostly GCs belonging to NGC\,4874) in this area. The difference is
$17\pm7$, which represents the number of GCs associated with SDSS\,J125935 brighter than
the GCLF turnover. For the total population, assuming a symmetric GCLF, we double this
number to obtain $34\pm14$~GCs. 
To estimate the specific frequency $S_N$ (number per unit $V$ luminosity; 
Harris \& van den Bergh 1981), we measure the galaxy magnitude
within the same elliptical aperture for consistency and find $\Iacs = 17.46\pm0.01$ mag.
The galaxy color is $\gIacs=1.072\pm0.018$ mag.  Both of these values are on the AB
system and are corrected for Galactic extinction.  Using empirical transformations from
Blakeslee et al.\ (2009, 2012), this color corresponds to 
$\gzacs \approx 1.26$~mag (typical of many dEs in the ACS Virgo Cluster Survey)
and $V{-}I \approx 1.10$~mag, where the latter value is on the standard Vega-based system.
To get the absolute $V$ magnitude, we subtract 0.42~mag from \Iacs\ to convert to the
Vega system, then subtract the distance modulus $\mM=35.0$, and finally add the estimated
$V{-}I$ to obtain $M_V=-16.86$~mag.  The specific frequency is then
$S_N = 6.1\pm2.6$ (the error includes an estimated 10\% uncertainty on the galaxy luminosity).
While this $S_N$ would be above average for a large galaxy, it is well within the
range for dEs of similar luminosity in the Virgo cluster (Peng et al.\ 2008).

Remarkably, SDSS\,J125935 has a heliocentric radial velocity
of $4193\pm154$~\kms, measured by Biviano \etal\ (1995). This is 
nearly 3000~\kms\ less than the velocity of $7176\pm3$~\kms\
for NGC\,4874 (Trager et al.\ 2008). 
According to Colless \& Dunn (1996), the
main component of the Coma cluster centered on NGC\,4874 has a mean velocity 
${\langle}v{\rangle} = 6853\pm54$~\kms\ and line-of-sight velocity dispersion
$\sigma_{\rm Coma} = 1082$~\kms.
Thus, SDSS\,J125935 has a relative velocity of $-2660\pm160$~\kms\ with respective to
the cluster mean, or $-(2.5\pm0.2)\sigma_{\rm Coma}$.
Numerical simulations indicate that dEs at such small clustercentric radii
and high relative velocities are likely to be on their first infall into the cluster
core (Smith \etal\ 2013, 2015). The same simulations show that dwarfs
that pass through the cluster centers can lose a large fraction, even the
majority, of their GC systems (see also Aguilar \& White 1986 for illustrations of how
similar encounters can result in asymmetric distributions of GC-like test masses). 

We suggest that SDSS\,J125935 is a dwarf elliptical with
a relatively rich GC system, similar to some dEs in Virgo, that has
recently fallen at high velocity into the core of the Coma cluster and is undergoing
stripping of its GC system. Unfortunately, it is presently unfeasible 
to measure spectroscopic velocities for the surrounding point sources to determine what
fraction belong to SDSS\,J125935. This would be another interesting system for detailed
dynamical modeling.

\section{Summary \& Conclusions}
\label{sec:conclusion}

We have studied the rich GC system of NGC\,4874, the cD galaxy in the core of the Coma
cluster of galaxies, using optical \hst\ \gacs\ and \Iacs\ imaging from the ACS/WFC and
near-IR \Hwfc\ imaging from WFC3/IR. The GC system of NGC\,4874 and 
the surrounding Coma core was previously studied in \gacs\ and \Iacs\ as part of the ACSCCS
(Peng \etal\ 2011), and we find excellent photometric agreement with that
study, but the exposure time of our \Iacs\ observations is more than
seven~times that of the ACSCCS imaging, giving a limiting magnitude more than a
magnitude fainter in this bandpass. Because we added the ACSCCS \gacs\ observations to
our own, the stacked \gacs\ image has a factor of two more exposure time than the ACSCCS
in the overlap region; tests show that our \gIacs\ color measurements have a factor of
two smaller errors than those from the ACSCCS. In addition, we include new deep F160W
observations, with an exposure time slightly longer than that of \Iacs, over the smaller
field of the WFC3/IR. 

Because the \IHacs\ color for old stellar populations measures red giant branch temperature,
it should be sensitive mainly to metallicity, while \gIacs\ also depends
on horizontal branch morphology and the location of the main sequence turnoff.
Over most of the luminosity range probed by our data, there exists
clear bimodality in the distribution of \gIacs\ colors of our selected GC candidates. 
This optical bimodality can be traced at least to $\Iacs=26$~mag, corresponding to 
$\MIacs=-9$~mag at the distance of Coma, or $M_V\approx-8$~mag for typical GCs.
From a Gaussian mixture modeling analysis as a function of magnitude, we find that at the
brightest magnitudes, the blue peak exhibits a very strong ``tilt'' towards redder colors, 
with a slope ${d(\gIacs)}/{d\Iacs} = -0.082\pm 0.020$ for $\Iacs<25$~mag. Based on the
empirical calibration of metallicity as a function of photometric color from 
the ACS Virgo Cluster Survey, this corresponds to a very steep mass-metallicity scaling
of $Z\propto M_{\rm GC}^{1.4\pm0.4}$ at these highest masses. 

The GMM analysis for the \IHhst\ color distribution is generally less robust
than for \gIacs, especially when the sample is further broken down by magnitude.
We therefore instead examined the variation in the overall mean \IHhst\ color in
the same magnitude range as for \gIacs.
Again for $\Iacs<25$~mag, we find a steep slope in the mean \IHhst\ color of
${d(\IHhst)}/{d\Iacs} = -0.093\pm0.013$. 
Although there is no empirical relation between \IHhst\ and metallicity
for GCs, the linear approximation to the relation between \IHhst\ and \gIacs\ gives
\hbox{${d(\IHhst)}/{d(\gIacs)}\approx1.1$}, which again implies $Z\propto M_{\rm GC}^{1.4}$.
{Thus, the mean metallicity scaling derived from the full \IHhst\ color 
  range is the same as that found from the blue component of the \gIacs\ color distribution.}
However, the color-magnitude tilt is not a simple linear relation, and if we extend
the linear fit another magnitude fainter to $\Iacs=26$~mag, then the best-fit slopes are
roughly a factor of three shallower, giving scalings of $Z\propto M_{\rm GC}^{0.5\pm0.2}$, 
consistent with the typical
scaling found by Mieske \etal\ (2010) over a similar mass range.

As a consequence of the tilted color-magnitude relations, the color distributions change
as a function of magnitude. Both the \gIacs\ and \IHhst\ distributions appear broad and
red, with no evidence for multiple peaks for the brightest GCs at
$\Iacs<23$~mag. Fainter than this, \gIacs\ is clearly bimodal, with the prominence of
the red peak decreasing at 
progressively fainter magnitudes. The bimodality is less evident in \IHacs, but the same
general trend occurs, with the histogram transitioning from a redward tilt to being
skewed towards the blue at fainter magnitudes. Because of the blue tilt at bright
magnitudes and increased measurement error at faint magnitudes, the bimodality is most
evident for $23<\Iacs<25$~mag, and we have compared the GMM bimodal
decompositions for \gIacs\ and \IHhst\ for the identical sample of GC candidates
over this magnitude range.  Once the four bluest
objects in \IHacs\ are excluded, the red:blue decompositions are consistent, with 
red fractions $f_2$ of $0.61\pm0.08$ for \gIacs\ and $0.54\pm0.13$ for \IHacs.

While the separation of the peaks in units of the peak dispersion is very clear
in \gIacs\ with $D=2.88\pm0.28$, it is less clear in \IHhst\ with $D=2.23\pm0.26$, even
though the separation in magnitudes is essentially identical. The reason for this is that
the blue peak is much narrower in \gIacs, with a dispersion $\sigma_1=0.067\pm0.014$~mag,
compared to $\sigma_1=0.098\pm0.013$~mag for \IHhst, a difference of nearly 50\%. For the
red peaks, the dispersions are $\sigma_2=0.105\pm0.014$~mag and $\sigma_2=0.126\pm0.017$~mag
for \gIacs\ and \IHhst, respectively. Previous studies of optical GC color
distributions (e.g., Peng \etal\ 2006, 2009; Harris \etal\ 2016) also found that the
blue peak was significantly narrower than the red peak; however, Peng \etal\ (2006) pointed out
that the dispersion in \feh\ was actually larger for the blue peak because of
the steeper variation in metallicity with color for the blue component of the GCs.
The differences in the blue and red color dispersions for \gIacs\ as compared to
\IHhst\ suggests that the colors follow different color-metallicity relations, despite
their nearly identical total range in color. In particular, the metallicity slopes at
blue and red colors must be more similar (i.e., weaker nonlinearity) for \IHhst\ than for
\gIacs. Consistent with this, we find that the variation in \IHhst\ with \gIacs\ is
nonlinear, with an inflected shape that can be described well by a cubic polynomial.

We have compared the radial distributions of the blue and red 
GCs over the wider ACS field of view. Consistent with previous studies, we find that the
blue GCs follow a more spatially extended radial profile than the red GCs. Interestingly,
for this field located in the dense central region of the rich Coma cluster of galaxies,
the broader extent of the blue GCs is at least partially the result of the GCs associated
with the fainter neighboring early-type cluster galaxies, whose GC systems are
predominantly blue, especially at large galactocentric radii.
This is consistent with the view that a significant fraction of the blue GCs in the
halos of massive galaxies are added through the accretion or stripping of lower 
luminosity satellite galaxies. 

Curiously, the center of the spatial distribution of the
GCs in this field is offset by $4\pm1$~kpc from the center of
NGC\,4874 itself. This offset does not appear to result from the superposition of the
GC population of any neighboring galaxy, but it is likely the signature of past dynamical
interaction. The most likely candidate for this is NGC\,4872, a bright SB0 galaxy 
24~kpc from the center of NGC\,4874 with a velocity difference of less than 20~\kms.
Although NGC\,4872 does not have a significant GC population of its own, the
4~kpc displacement in the centroid of the NGC\,4874 GC system lies along the line
towards NGC\,4872. We have also discussed the asymmetry of the GC system of the
dE galaxy SDSS\,J125935, which is projected 42~kpc from NGC\,4874, but has a
relative velocity of $-2983$~\kms\ with respect to the cD, and $-2660$~\kms\ with respect
to the cluster mean. The dE has a specific frequency $S_N = 6.1\pm2.6$.
The likelihood of the asymmetry in its GCs occurring by chance is $\sim1$\%. 
We suggest that this dE is on an initial high-velocity infall into
the cluster core and its GC system is in the process of being stripped.

Interestingly, based on stellar absorption line indices, Trager \etal\ (2008) concluded
that NGC\,4874 and neighboring early-type galaxies showed evidence for an
intermediate-age stellar population component, which would imply a significant star
formation event several billion years~ago. For now, it remains a matter of speculation
whether this proposed star formation event in the relatively recent past is associated
with the spatial offset of the NGC\,4874 GCs. It is also unknown whether or not such
an event may have produced any significant population of intermediate-age GCs. If so,
one would expect the color-metallicity and color-color relations in this field to differ
from those in massive galaxies with exclusively old GC populations, as predicted from 
stellar population models (Yoon \& Chung 2009). Usher \etal\ (2015)
have shown that the color-metallicity relations do indeed vary among early-type galaxies,
and that this variation appears to correlate with galaxy luminosity and color; further
work is needed to understand the detailed causes of these variations.

We are currently carrying out an optical-NIR photometric study of GCs in a much larger
set of 16 early-type galaxies in the Fornax and Virgo clusters by cross-matching our
\hst\ WFC3/IR data (Jensen \etal\ 2015) with the published F475W and F850LP catalogs from
the ACS Fornax and Virgo Cluster Surveys (Jord{\'a}n \etal\ 2009, 2015).
Because these galaxies cover a large range in luminosity and color, this sample will
shed light on whether optical/NIR color-color relations show variations with galaxy type
similar to those found by Usher \etal\ (2015) for the relation between optical color and
metallicity estimated from the CaT index, as well as illuminating differences in
the ways that different broadband colors trace the underlying metallicity. 
Unfortunately, existing samples of spectroscopically estimated metallicities for massive
early-type galaxies are of inhomogeneous quality, tend to be based on a small number of metal
absorption line indices, and often have large uncertainties in excess of $0.5$~dex. 
A large sample of uniformly high-quality spectroscopic metallicities
($\sigma_{\feh}\sim0.1$ dex) and ages determined over a broad spectral range
for hundreds of GCs spanning the full color range in a nearby cD galaxy 
(which likely combines GCs from a diverse mix of other cluster galaxies)
would be an invaluable resource for the community. 
Such a sample would allow us to calibrate empirically the detailed
forms of the color-metallicity relations from the UV to the NIR, and thus constrain the
enrichment histories of more distant galaxies from photometric studies alone;
it would also enable crucial tests of the stellar population models.
The evolutionary histories of massive galaxies and their surrounding environments are encoded
in the properties of the ancient systems of GCs that surround them; decoding these
histories remains a major ongoing archaeological effort in extragalactic astronomy.

\acknowledgements 
Support for this work was provided by the National Research Foundation of Korea
to the Center for Galaxy Evolution Research (CGER). H.C.~thanks the National Research
Council of Canada's Herzberg Astronomy \& Astrophysics for hospitality during several visits.
J.P.B.~thanks the CGER at Yonsei University for hospitality on numerous occasions.
We thank Pat~C{\^o}t{\'e}, Ruben Sanchez-Janssen, Laura Sales, and Rory Smith for helpful conversations.
This research has made use of the NASA/IPAC Extragalactic Database
(NED) which is operated by the Jet Propulsion Laboratory, California Institute
of Technology, under contract with the National Aeronautics and Space
Administration.

{\it Facility:} \facility{\HST\ (WFC3/IR, ACS/WFC)}

\clearpage

\begin{deluxetable*}{crrrrrrrrr}
\tablecolumns{10}
\tablewidth{0pc}
\tabletypesize{\footnotesize}
\tablecaption{The GMM Analysis Results for the Color Distributions Shown in Figures~\ref{fig:hist1}, \ref{fig:hist2}, and \ref{fig:hist3} (full sample)}
\tablehead{
\colhead{\Iacs}     & \colhead{N$_{\rm tot}$} & \colhead{$\mu_{1}$}  & \colhead{$\sigma_{1}$}  & \colhead{$\mu_{2}$} & \colhead{$\sigma_{2}$}  &
\colhead{$f_{2}$}   & \colhead{{\it D}}       & \colhead{{\it kurt}} & \colhead{$p(\chi^2)$}   \\
\colhead{(1)} & \colhead{(2)} & \colhead{(3)} & \colhead{(4)}   & \colhead{(5)}  & \colhead{(6)}  &
\colhead{(7)} & \colhead{(8)} & \colhead{(9)} & \colhead{(10)} }
\startdata
\sidehead{{\gIacs} of ACS F814W and F475W matched samples}
21.5--23.0 & 26 & 1.065$\pm$0.030 & 0.048$\pm$0.017 & 1.219$\pm$0.028 & 0.053$\pm$0.016 & 0.566$\pm$0.167 & 3.07$\pm$0.72 & $-$1.105 & 0.566  \\
23.0--24.0 & 164 & 0.922$\pm$0.035 & 0.078$\pm$0.021 & 1.154$\pm$0.026 & 0.091$\pm$0.015 & 0.634$\pm$0.126 & 2.74$\pm$0.40 & $-$0.756 & 0.003  \\
24.0--25.0 & 775 & 0.866$\pm$0.011 & 0.068$\pm$0.010 & 1.112$\pm$0.022 & 0.127$\pm$0.012 & 0.627$\pm$0.071 & 2.42$\pm$0.26 & $-$0.736 & $<$0.001 \\
25.0--26.0 & 1798 & 0.839$\pm$0.010 & 0.071$\pm$0.014 & 1.090$\pm$0.034 & 0.190$\pm$0.017 & 0.702$\pm$0.092 & 1.75$\pm$0.28 & $-$0.557 & $<$0.010  \\
23.0--25.0\tablenotemark{a} & 939 & 0.873$\pm$0.010 & 0.071$\pm$0.008 & 1.123$\pm$0.017 & 0.119$\pm$0.010 & 0.624$\pm$0.055 & 2.55$\pm$0.21 & $-$0.775 & $<$0.001 \\
\sidehead{{\IHacs} of ACS F814W and WFC3/IR F160W matched samples}
21.5--23.0 & 12 & 0.638$\pm$0.036 & 0.102$\pm$0.028 & 0.914$\pm$0.050 & 0.026$\pm$0.019 & 0.161$\pm$0.129 & 3.71$\pm$1.23 & $-$0.854 & 0.275 \\
23.0--24.0 & 74 & 0.400$\pm$0.032 & 0.071$\pm$0.018 & 0.633$\pm$0.032 & 0.088$\pm$0.016 & 0.603$\pm$0.119 & 2.92$\pm$0.49 & $-$0.899 & 0.088 \\
24.0--25.0 & 319 & 0.397$\pm$0.044 & 0.138$\pm$0.024 & 0.675$\pm$0.078 & 0.110$\pm$0.031 & 0.271$\pm$0.184 & 2.24$\pm$0.41 & $-$0.483 & 0.050 \\
25.0--26.0 & 643 & 0.419$\pm$0.167 & 0.202$\pm$0.040 & 0.769$\pm$0.111 & 0.138$\pm$0.054 & 0.152$\pm$0.240 & 2.02$\pm$0.71 & $-$0.111 & 0.190 \\
23.0--25.0\tablenotemark{a} & 393 & 0.399$\pm$0.035 & 0.133$\pm$0.020 & 0.660$\pm$0.052 & 0.108$\pm$0.020 & 0.330$\pm$0.144 & 2.17$\pm$0.32 & $-$0.499 & 0.030 \\
\enddata
\tablecomments{Column lists: (1) \Iacs\ AB magnitude range; (2) total number of objects in the analyzed sample; (3) mean and uncertainty of the first mode in the double Gaussian mixture model; (4) standard deviation and uncertainty of the first mode; (5) mean and uncertainty of the second mode; (6) standard deviation and uncertainty of the second mode; (7) fraction of objects assigned to the second component of the double Gaussian mixture model; (8) separation between the peaks relative to their Gaussian $\sigma$; (9) kurtosis of the distribution; (10) likelihood that the sample was drawn from a single Gaussian distribution.}
\tablenotetext{a}{The results for the color distributions over the $23.0 < \Iacs < 25.0$ magnitude range shown in Figure~\ref{fig:hist3}.}
\label{tab:GMM1}
\end{deluxetable*}

\begin{deluxetable*}{crrrrrrrrr}
\tablecolumns{10}
\tablewidth{0pc}
\tabletypesize{\footnotesize}
\tablecaption{\small The GMM Analysis Results for the Color Distributions Shown in Figure~\ref{fig:hist4} (matched subsample)}
\tablehead{
\colhead{Case}    & \colhead{N$_{\rm tot}$} & \colhead{$\mu_{1}$}  & \colhead{$\sigma_{1}$} & \colhead{$\mu_{2}$} & \colhead{$\sigma_{2}$} &
\colhead{$f_{2}$} & \colhead{{\it D}}         & \colhead{{\it kurt}} & \colhead{$p(\chi^2)$}\\
\colhead{(1)} & \colhead{(2)} & \colhead{(3)} & \colhead{(4)}  & \colhead{(5)}  & \colhead{(6)}  &
\colhead{(7)} & \colhead{(8)} & \colhead{(9)} & \colhead{(10)} 
}
\startdata
\sidehead{{\gIacs} of ACS F814W $+$ F475W $+$ WFC3/IR F160W matched samples}
$\sigma_{1}\neq\sigma_{2}$ & 392 & 0.884$\pm$0.016 & 0.067$\pm$0.015 & 1.139$\pm$0.026 & 0.106$\pm$0.016 & 0.604$\pm$0.083 & 2.89$\pm$0.32 & $-$0.986 & $<$0.001 \\
$\sigma_{1}=\sigma_{2}$ & 392 & 0.911$\pm$0.009 & 0.087$\pm$0.004 & 1.168$\pm$0.009 & 0.087$\pm$0.004 & 0.493$\pm$0.031 & 2.94$\pm$0.20 & $-$0.986 & $<$0.001 \\
\sidehead{{\IHacs} of ACS F814W $+$ F475W $+$ WFC3/IR F160W matched samples}
$\sigma_{1}\neq\sigma_{2}$ & 392 & 0.399$\pm$0.031 & 0.133$\pm$0.018 & 0.663$\pm$0.047 & 0.107$\pm$0.018 & 0.324$\pm$0.128 & 2.18$\pm$0.29 & $-$0.501 & 0.034 \\
$\sigma_{1}=\sigma_{2}$ & 392 & 0.373$\pm$0.015 & 0.121$\pm$0.008 & 0.629$\pm$0.019 & 0.121$\pm$0.008 & 0.436$\pm$0.054 & 2.12$\pm$0.28 & $-$0.501 & 0.007 \\
\enddata
\tablecomments{Column (1) shows whether the double Gaussian mixture is for the heteroscedastic or homoscedastic case. Columns (2)-(10) are the same as in Table~\ref{tab:GMM1}.}
\label{tab:GMM2}
\end{deluxetable*}

\begin{deluxetable*}{rrrrrrrrrr}
\tablecolumns{10}
\tablewidth{0pc}
\tabletypesize{\footnotesize}
\tablecaption{\small Same as Table~\ref{tab:GMM2} but in different color ranges.}
\tablehead{
\colhead{\IHacs} & \colhead{N$_{\rm tot}$} & \colhead{$\mu_{1}$}  & \colhead{$\sigma_{1}$}  & \colhead{$\mu_{2}$} & \colhead{$\sigma_{2}$} &
\colhead{$f_{2}$}     & \colhead{{\it D}} & \colhead{{\it kurt}} & \colhead{$p(\chi^2)$}  \\
\colhead{(1)} & \colhead{(2)} & \colhead{(3)} & \colhead{(4)}   & \colhead{(5)}  & \colhead{(6)}  &
\colhead{(7)} & \colhead{(8)} & \colhead{(9)} & \colhead{(10)} 
}
\startdata
\sidehead{{\gIacs} of ACS F814W $+$ F475W $+$ WFC3/IR F160W matched samples}
$-0.022$ to $0.877$ & 392 & 0.884$\pm$0.016 & 0.067$\pm$0.015 & 1.139$\pm$0.026 & 0.106$\pm$0.016 & 0.604$\pm$0.083 & 2.89$\pm$0.32 & $-$0.986 & $<$0.001 \\
$ 0.059$ to $0.877$ & 390 & 0.885$\pm$0.018 & 0.067$\pm$0.016 & 1.140$\pm$0.026 & 0.105$\pm$0.016 & 0.604$\pm$0.087 & 2.88$\pm$0.30 & $-$0.980 & $<$0.001 \\
$ 0.133$ to $0.877$ & 388 & 0.885$\pm$0.016 & 0.067$\pm$0.014 & 1.139$\pm$0.023 & 0.105$\pm$0.014 & 0.609$\pm$0.082 & 2.88$\pm$0.28 & $-$0.978 & $<$0.001 \\
\sidehead{{\IHacs} of ACS F814W $+$ F475W $+$ WFC3/IR F160W matched samples}
$-0.022$ to $0.877$ & 392 & 0.399$\pm$0.031 & 0.133$\pm$0.018 & 0.663$\pm$0.047 & 0.107$\pm$0.018 & 0.324$\pm$0.128 & 2.18$\pm$0.29 & $-$0.501 & 0.034 \\
$ 0.059$ to $0.877$ & 390 & 0.368$\pm$0.033 & 0.111$\pm$0.015 & 0.627$\pm$0.042 & 0.119$\pm$0.017 & 0.461$\pm$0.132 & 2.25$\pm$0.26 & $-$0.667 & 0.002 \\
$ 0.133$ to $0.877$ & 388 & 0.353$\pm$0.031 & 0.098$\pm$0.013 & 0.606$\pm$0.041 & 0.126$\pm$0.017 & 0.538$\pm$0.126 & 2.23$\pm$0.26 & $-$0.727 & $<$0.001 \\
\enddata
\tablecomments{Column (1) shows minimum and maximum values of {\IHacs} color for the objects in the sample (the {\gIacs} ranges are not modified). Columns (2)-(10) are the same as in Table~\ref{tab:GMM1}.}
\label{tab:GMM3}
\end{deluxetable*}

\end{document}